\documentclass[preprint, amsmath,amssymb,onecolumn]{revtex4-2}
\usepackage{setspace}
\usepackage{graphicx}
\usepackage{dcolumn}
\usepackage{bm}
\usepackage[shortlabels]{enumitem}

\usepackage{lineno,hyperref}
\usepackage{xcolor}
\usepackage{physics}
\usepackage[version=4]{mhchem}
\modulolinenumbers[5]
\hypersetup{colorlinks = true,linkcolor = blue,anchorcolor =red,citecolor = blue,filecolor = red,urlcolor = black,
            pdfauthor=author}

\providecommand\Ra{\mathit{Ra}}

\newcommand\Csat{C_\mathit{sat}}
\providecommand\dd{\mathrm{d}}

\begin{document}



\title{Diffusive and convective dissolution of carbon dioxide in a vertical cylindrical cell}

\author{Dani\"{e}l P. Faasen}
\thanks{These two authors contributed equally.}
\email{d.p.faasen@utwente.nl}
\author{Farzan Sepahi}%
\thanks{These two authors contributed equally.}
\email{f.sepahi@utwente.nl}
\author{Dominik Krug}%
\email{d.j.krug@utwente.nl}
\author{Roberto Verzicco}%
\email{r.verzicco@utwente.nl}
\author{Pablo Pe\~{n}as}%
\author{Detlef Lohse}%
\email{d.lohse@utwente.nl}
\author{Devaraj van der Meer}%
\email{d.vandermeer@utwente.nl}

\affiliation{%
Physics of Fluids Group, Faculty of Science and Technology, University of Twente, P.O. Box 217, 7500 AE Enschede, The Netherlands\\
}%

\date{\today}

\begin{abstract}

The dissolution and subsequent mass transfer of carbon dioxide gas into liquid barriers plays a vital role in many environmental and industrial applications. In this work, we study the downward dissolution and propagation dynamics of CO$_2$ into a vertical water barrier confined to a narrow vertical glass cylinder, using both experiments and direct numerical simulations. Initially, the dissolution of CO$_2$ results in the formation of a CO$_2$-rich water layer, which is denser in comparison to pure water, at the top gas-liquid interface. Continued dissolution of CO$_2$ into the water barrier results in the layer becoming gravitationally unstable, leading to the onset of buoyancy driven convection and, consequently, the shedding of a buoyant plume. By adding sodium fluorescein, a pH-sensitive fluorophore, we directly visualise the dissolution and propagation of the CO$_2$ across the liquid barrier. Tracking the CO$_2$ front propagation in time results in the discovery of two distinct transport regimes, a purely diffusive regime and an enhanced diffusive regime. Using direct numerical simulations, we are able to successfully explain the propagation dynamics of these two transport regimes in this laterally strongly confined geometry, namely by disentangling the contributions of diffusion and convection to the propagation of the CO$_2$ front.  

\end{abstract}

\maketitle
\section{Introduction}

The dissolution and subsequent mass transfer of carbon dioxide gas into liquid barriers plays a vital role in many environmental and industrial applications. In microfluidics for example, Taylor flow, a segmented flow of alternating gas and liquid plugs, is utilised in microreactor designs to increase heat and mass transfer rates, resulting in higher reactor performance \cite{Cao2021, Bagemihl2022, Abolhasani2014}. On a larger scale, carbon capture and sequestration is often based on injecting CO$_2$ into deep saline aquifers, trapping the CO$_2$ between a layer of cap rock and a liquid reservoir, which results in the long term, stable storage of CO$_2$ in the aquifer \cite{Hassanzadeh2007,Emami2015,DePaoli2021}. 

Once the carbon dioxide starts to dissolve into the water layer, a CO$_2$-rich water layer forms at the interface, which is denser in comparison to pure water. While initially stable, the continued dissolution of CO$_2$ into the water layer results in the CO$_2$-rich fluid layer becoming gravitationally unstable, leading to the onset of buoyancy driven convection and the formation of a buoyant plume, which greatly enhances the mass transfer of CO$_2$ in the water layer \cite{Backhaus2011,Loodts2014}. Furthermore, density driven convection can also occur as a result of buoyancy generating chemical reactions \cite{Rogers2005,Loodts2018,Gu2018,Sepahi2022}, droplet dissolution \cite{Dietrich2016, Chong2020} and bubble growth \cite{Enriquez2014,Soto2019}.  

In literature, studies investigating the dissolution and density driven convection in the CO$_2$--water system have reported between two and four distinct transport regimes \cite{Moghaddam2012,Gholami2015,Tang2019}. These regimes are vaguely defined by their assumed dominant driving mechanism and thus referred to as, for example, "purely diffusive", "early convective", or "late convective" \cite{Moghaddam2012}. Moreover, in the regimes where convection is contributing to the mass transport, apparent diffusive behaviour is observed, albeit with a much higher effective diffusion coefficient. Depending on the experimental conditions, this effective diffusion coefficient can be several orders of magnitude bigger in comparison to the expected diffusive counterpart under similar experimental conditions \cite{Du2019,Karimaie2017,Farajzadeh2009,Yang2019,Zhao2018}. However, little explanation has been given as to what drives the different observed regimes, the transitions between the regimes and why the system still appears to behave in a diffusive manner. 

This is precisely the focus of our work. We study the dissolution and downward propagation of CO$_2$ into a vertical water barrier confined to a narrow cylindrical cell either above a trapped air bubble, an alkane layer or directly on top of a solid silicon plate, as shown in figure \ref{fig_schematic}. We replace the ambient air atmosphere with a CO$_2$ atmosphere at the same pressure and by adding sodium fluorescein, a pH-sensitive fluorophore, to the liquid barrier, we can directly visualise the propagation of CO$_2$ \cite{Doughty2010,Robertson2013,Yewalkar2019}. We compare the experimental results to those obtained by 3D direct numerical simulations, in order to elucidate the relevant transport mechanisms.

In a nutshell, our aim is to investigate the mass transport mechanisms in a laterally strongly confined system after the dissolution of CO$_2$ into a liquid barrier. We will identify two different regimes, namely: a purely diffusive regime and an enhanced diffusive regime. The direct numerical simulations allow us to disentangle the contributions of the buoyancy driven convection and diffusion towards the front propagation velocity. We will show that the onset of convection leads to a distortion of the propagation front surface, resulting in the increase of the concentration gradients which in turn leads to enhanced diffusive fluxes. The overall behaviour remains diffusive however, although with an increased diffusion coefficient. The diffusive propagation acts to flatten the interfacial area, which over time leads to an equilibrium with the convective bulging of the front, after which the front propagates at an almost constant velocity. 

This paper is organised as follows. The experimental setup and procedure are described in Sec. \ref{sec:Experimental}.  Section \ref{sec:Experimentalobservations} presents the results of our visualisation experiments. In Sec. \ref{sec:Intensityprofiles} the amount of CO$_2$ in the liquid barrier over time is investigated by first obtaining the intensity profiles from the visualisation experiments and subsequently converting them to concentration profiles. In Sec. \ref{sec:Frontpropagation} we study the front propagation dynamics of the CO$_2$ layer and identify two distinct propagation regimes. In Sec. \ref{sec:NumericalModel} we provide the details on the numerical model we us to study the physics behind the front propagation dynamics and provide a comparison between the numerical model and the experiments. The paper ends with a summary of the main findings and an outlook in Sec. \ref{sec:conclusions}.

\section{Experimental procedure}
\label{sec:Experimental}

\begin{figure}
	\centering
	\includegraphics[width=0.75\textwidth]{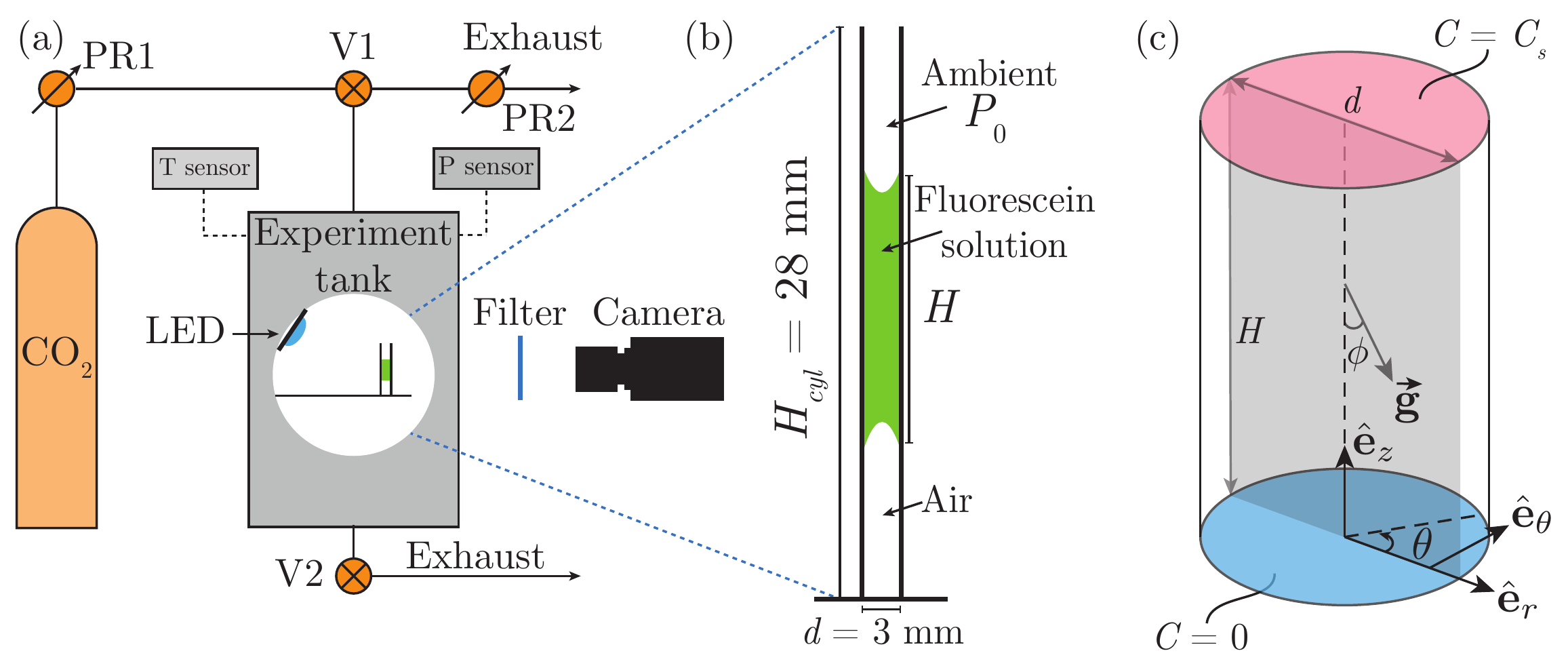}
	\caption{
	(a) Schematic overview of the experimental setup. (b) Sketch of the cylinder containing the liquid--air setup. The cylinder is placed inside the experimental chamber, which is subsequently flushed with CO$_2$ gas. (c) Schematic overview of the numerical setup.}
	\label{fig_schematic}
\end{figure}

A schematic overview of our experimental setup is shown in figure \ref{fig_schematic}a. The experiments are conducted inside a sealed chamber which can be flushed with CO$_2$ gas. The inlet pressure is fixed to 1.0 bar using the pressure regulator PR1, whereas pressure regulator PR2 prevents over-pressurisation of the experimental tank. A more detailed description of the experimental chamber and pressure control system can be found elsewhere \cite{Enriquez2013}.

A single borosilicate glass (Duran) cylinder (28 mm in length, inner diameter $d$ = 3.0 mm, outer diameter of 5.0 mm) is attached on one end to a silicon wafer plate using Loctite 4305 (Farnell), in an almost perfectly vertical manner, while the other end is left open. Before use, the cylinder is rinsed using ethanol (Boom, technical grade) followed by Milli-Q water (resistivity = 18.2 M$\Omega$ cm) and finally dried in a nitrogen stream. 

We prepare the cylinder in one of three different configurations, a liquid--solid, liquid--liquid, or a liquid--air configuration, as depicted in figure \ref{fig_schematic}b. In all configurations, a layer of a $10^{-4}$ M aqueous fluoroscein solution is injected into the cylinder, at a volume of $V$ = 120 $\mu$L (or $H$ $\approx$ 18 mm), which acts as the liquid barrier. This solution is freshly prepared prior to the experiments by adding sodium fluorescein salt (Fisher Scientific, general purpose grade) to Milli-Q water. Fluoroscein is a well known fluorophore often used in biological application, with its main absorbance peak at 490 nm and main emission peak at 513 nm \cite{Yewalkar2019,Doughty2010,Robertson2013}. More importantly, the emission intensity of fluorescein has a (non-linear) dependency on the pH level of the liquid, allowing us to follow the dissolution and propagation of the CO$_2$ in the liquid. Furthermore, the presence of the sodium fluorescein in the barrier does not affect the diffusive and convective behaviour of the CO$_2$, as we have $\sim$ 4.4 $\mu$g sodium fluorescein in the 120 $\mu$L barrier to achieve the desired concentration.

For the liquid--air configuration, the fluorescein layer is placed in the cylinder such that a bubble of arbitrary height spanning the entire width of the cylinder, is trapped underneath the liquid. The liquid barrier remains in place due to a stable balance between the surface tension of the liquid--air interface, the weight of the liquid barrier, and the differences in gas pressures. Since the surface tensions of CO$_2$ and air above water are almost identical, the force balance persists throughout our experiments \cite{Chun1995}. For the liquid--solid configuration, the fluoroscein solution is injected into the cylinder such that no air is trapped between the liquid barrier and the silicon wafer plate. During this process, special attention is paid to ensure no small bubbles are entrained at the liquid--solid interface. Finally, for the liquid--liquid interface, 60 $\mu$L $n$-hexadecane (VWR, 99$\%$ purity) is injected first into the cylinder. On top of this liquid layer, the fluorescein solution is carefully injected, again to prevent the entrapment of bubbles. Despite the density of $n$-hexadecane being lower compared to the density of water, this configuration remains stable during our experiments, again due to a stable balance between the surface tensions at the interface and the weight of the top liquid column. 

Inside the tank, a LED (Thorlabs, $\lambda _{centre}$ = 470 nm) is located to illuminate the cylinder, while the pressure and temperature sensors in the chamber record the pressure $P_0$ and temperature $T$ in time respectively (1 acquisition per second). The average temperature during our experiments is determined to be $T = 22.3 \pm 0.4$ $^\circ$C. We use a Nikon D850 camera in silent interval timer shooting mode (1 fps) in combination with a Zeiss Makro Planar T 100 mm lens to achieve a mean optical resolution of 10.4 $\mu$m/pixel. In the optical path between the cylinder and the camera, a bandpass filter (Thorlabs,  $\lambda _{centre}$ = 530 nm, BW = 43 nm) is located to block out the LED light.

After preparation, the cylinder is placed inside the experimental chamber. The inlet pressure is fixed to 1.0 bar using PR1 while valve V1 remains closed. Valve V2 is opened to allow the experimental tank to be flushed during the flushing stage. The LED inside the chamber is turned on and 5 seconds later the  interval timer shooting mode on the camera is activated. 15 seconds after camera activation, the pressure and temperature sensor data starts being recorded. Finally, 35 seconds after turning on the LED inside the chamber, valve V1 is opened and the system is flushed with CO$_2$ gas in order to fully replace the ambient air inside the tank. The time at the start of the flushing stage is $t$ = 0 seconds and marks the start of the "experiment" stage. After flushing for 60 seconds, valves V1 and V2 are closed in quick succession, with the former being closed first to prevent pressurisation of the experimental tank. At the end of the experiment (typically at $t$ = 15 minutes), the experimental tank is opened and flushed using a nitrogen spray gun to prepare the experimental chamber for the next experiment.

Based on the aforementioned experimental conditions, we can calculate the relevant dimensionless numbers. Our Schmidt number is found to be Sc =$\nu / D$ =  515. Since the maximum CO$_2$ concentration difference in the barrier is $\Delta C = \Csat$, we find our maximum Rayleigh number to be:

\begin{equation}
\Ra_H   \equiv \frac{\beta \Csat g H^3}{\nu D}, 
\end{equation}

\noindent where $g$ is the acceleration due to gravity, $\beta$ = (8.2 $\pm$ 0.03) cm$^3$/mol the solutal expansion coefficient of CO$_2$, the saturation concentration $\Csat = k_H P_0$, with $k_H$ = (3.53 $\pm$ 0.04) $\times$ 10$^{-4}$ mol/m$^3$Pa and $P_0$ = 1.0 bar, $D$ = (1.85 $\pm$ 0.02) $\times$ 10$^{-9}$ m$^2$/s the diffusion coefficient of CO$_2$ in water, $H$ = 17.6 $\pm$ 0.35 mm the height of the liquid barrier, and $\nu$ = 9.5 $\times$ 10$^{-7}$ m$^2$/s the kinematic viscosity of water \cite{Sander2015,Tamimi1994,Loodts2014}. We obtain Ra$_H$ $\approx$  (8.8 $\pm$ 0.5) $\times$ 10$^{6}$ , which is well above the critical Rayleigh number, $Ra_{H,c} = 1.29\times 10^{6}$, based on the minimal aspect ratio  ($\Gamma_{max} = d/H = 0.17$) of our experimental setup \cite{Ahlers2022}.

\section{Experimental observations}
\label{sec:Experimentalobservations}

\begin{figure}
	\centering
	\includegraphics[width=0.9\textwidth]{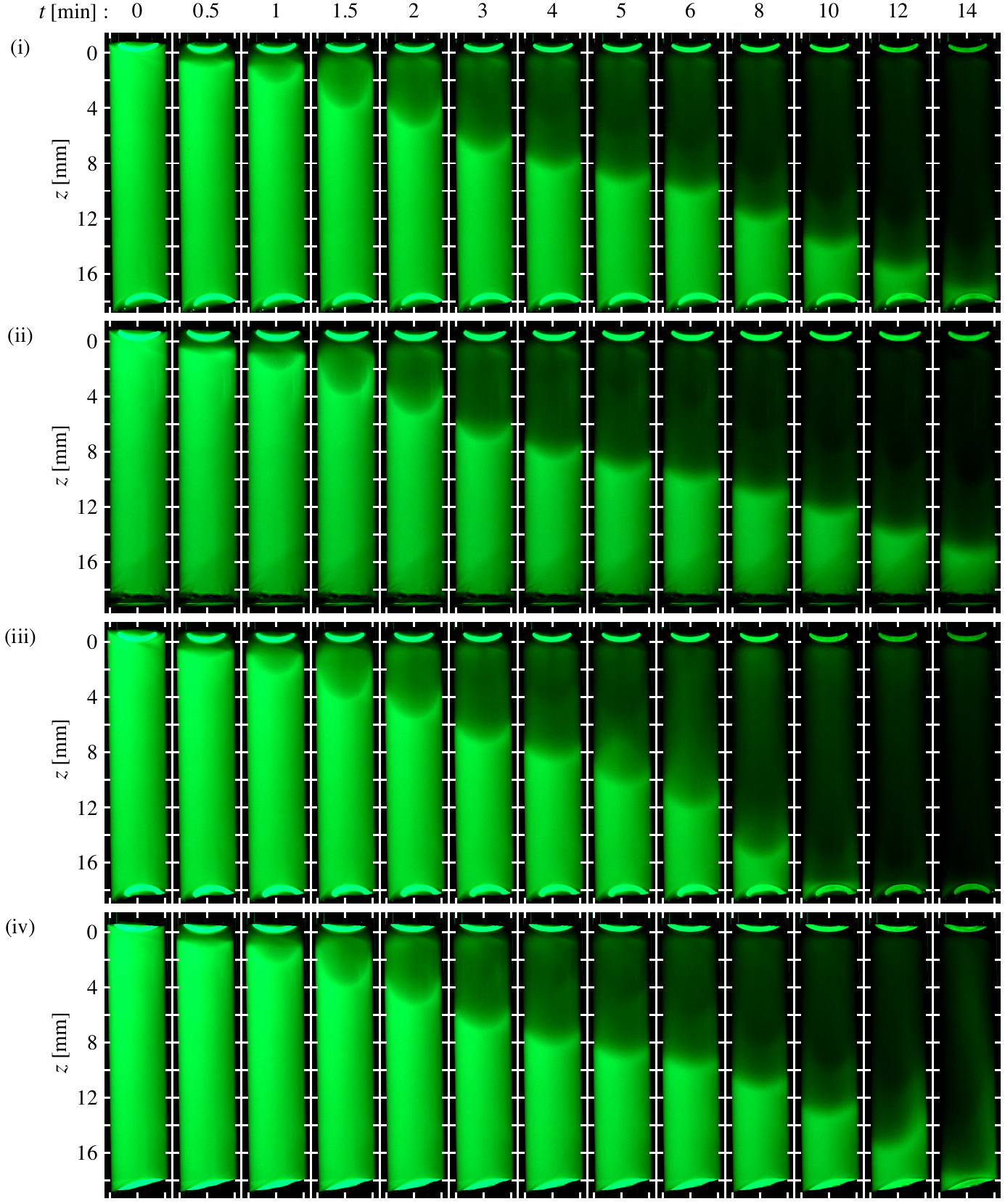}
	\caption{Fluorescence images of the initial dissolution process of CO$_2$ in a vertical liquid column within a cylindrical cell.  The fluorescence intensity decays with pH or increasing CO$_2$ concentration. At $t= 0$, the upper interface  is exposed to a CO$_2$ gas ambient. Subsequently, a CO$_2$-containing layer (dark region) propagates downwards. The bottom liquid interface for (i, iii, iv)  is liquid--air; for (ii) it is liquid--solid. The type of boundary has no impact on the propagation dynamics of the CO$_2$  front. In (i) and (ii) the  front propagates  axisymmetrically  throughout the entire water depth. In (iv), axisymmetry is broken at $t \approx 12$ min  with the shedding of a lateral buoyant upwelling plume. A similar symmetry-breaking upwelling plume  occurs in (iii) at  $t \approx 5$ min, yet the plume appears visually centred due to the planar visualisation of the 3D system. Such an event causes the front to accelerate towards a higher velocity. Coordinate $z$ denotes the depth from the apex of the top meniscus; the horizontal ticks are 2 mm apart.}
	\label{fig_snapshots}
\end{figure}

We begin by analysing a series of liquid--air experiments, (i), (iii), and (iv) in figure \ref{fig_snapshots}, and a liquid--solid experiment, (ii) in figure \ref{fig_snapshots}. Snapshots of additional experiments can be found in the supporting material in figures S1 and S2. As the CO$_2$ starts dissolving into the liquid barrier, the pH of the CO$_2$ imbued liquid starts to decrease. Since fluorescein is a pH sensitive fluorophore in the range 5 $\lesssim$ pH $\lesssim$ 10, the emission intensity of the fluorescein dye starts to decrease, resulting in a colour change of the dye from bright green to black in the images \cite{Pande2020}. 

At the beginning of the experiments, $t$ = 0 s, we start replacing the air atmosphere with a CO$_2$ atmosphere. Almost immediately, CO$_2$ starts dissolving into the liquid barrier, forming a CO$_2$-rich water layer just below the gas--liquid interface. While initially stable, the continued dissolution of CO$_2$ into the water barrier results in the layer becoming gravitationally unstable, as the CO$_2$-rich water is denser in comparison to the pure water underneath. Once this happens, a convective plume is shed from the CO$_2$-rich boundary layer which starts propagating downwards into the liquid barrier. In the experiments shown, the shedding of the buoyant plume occurs around t $\sim$ 1 minute.

After shedding the convective plume, differences in the CO$_2$ front propagation dynamics can be observed. In (i) and (ii) the front appears to propagate axisymmetrically throughout the water barrier. As a result, the CO$_2$ front does not quite reach the lower liquid--gas, (i), or liquid--solid, (ii), interface. In contrast, the apparent axisymmetry observed in (i) and (ii) is seemingly broken in (iii) and (iv) at $t$ = 5 minutes and $t$ = 12 minutes, respectively. In (iv) the shedding of a lateral buoyant upwelling plume can be observed at the right side of the cylinder at the interface of the denser CO$_2$-rich liquid and the pure bulk liquid. In (iii), the plume appears visually centred due to the planar visualisation of the 3D system. Regardless, in both cases the CO$_2$ front accelerates towards a higher velocity, leading to the front reaching the bottom liquid--air interface around t $\sim$ 10 min in experiment (iii) and  t $\sim$ 15 min in experiment (iv). It therefore appears that the type of bottom boundary does not affect the CO$_2$ propagation dynamics, but the occurrence of the shedding of a secondary plume does.

\begin{figure}
	\centering
	\includegraphics[width=0.5\textwidth]{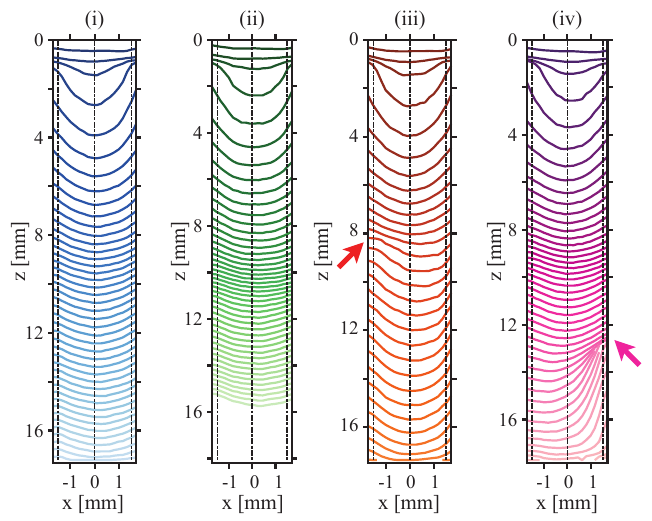}
	\caption{(a) Propagation of the projected front surface for experiments (i--iv) shown in  figure \ref{fig_snapshots} during $0<t<15$ min. The front is defined as  the iso-concentration contour $z_f = z_f(x, t)$ corresponding to a normalised intensity $I^*(x, \ z, \ t)$ value of 60 \%. The front is computed from the vertical intensity profiles computed across 20 equispaced positions in $x$.
	The time step between contour lines is $\Delta t = 20$ s. In (iii) and (iv), the arrows denote the front positions at which lateral plume shedding occurs, i.e.,  when the axisymmetric propagation is broken.}  		
	\label{fig_front2d}
\end{figure}

To further study the propagation dynamics, we track the progression of the CO$_2$ front in the liquid barrier over time. We define 20 equispaced bins along the cylinders diameter for which the vertical intensity profiles are calculated. The obtained intensity profiles are normalised with respect to the intensity profiles at $t$ = 0s, in order to account for variations in illumination. The CO$_2$ front is defined as the iso-concentration contour $z_f = z_f (x,t)$ corresponding to a normalised intensity $I^*(x, \ z, \ t)$ value of 60$\%$. For experiments (i-iv), the obtained projected front surfaces are shown in figure \ref{fig_front2d}. The time step between the contour lines is $\Delta t$ = 20s. The front profiles of the additional experiments can be found in the supporting material in figures S3 and S4.

The shown front contours emphasise that the initial behaviour for the four experiments is very similar. After the shedding of the buoyant plume by the diffusive boundary layer, the front initially rapidly accelerates, and then slows down again as time progresses. As mentioned, in experiments (i) and (ii), the front reaches a stable velocity, indicated by the front contours becoming equidistant in space. For experiments (iii) and (iv), the arrows indicate the depth at which we observe the shedding of a lateral buoyant plume. After this event, the spatial distance between the lines increases again, indicating the acceleration of the front to a higher velocity. This is very similar to the initial shedding event observed from the diffusive boundary layer. The secondary plume shedding is not observed in all cases and is therefore likely related to uncontrolled noise in the experiments, such as small, local deviations in CO$_2$ concentration or the small inclination of the cylinder with respect to the base plate.

\section{Intensity and concentration profiles}
\label{sec:Intensityprofiles}

We continue our analysis by investigating the amount of CO$_2$ absorbed in the liquid barrier over time. In order to do so, we first have to obtain the intensity profiles of the fluorescein solution in the barrier and convert these to the corresponding CO$_2$ concentration profiles. Therefore, we start by computing the horizontally averaged intensities $G(z, \ t)$ which are then a function of depth $z$ and time $t$ only. We normalise the obtained intensities with respect to the initial intensity profile in order to account for spatial inhomogeneity of the LED lighting, defined as the normalised intensity $I  = G(z, \ t)/G(z, \ 0)$. Additionally, we correct these intensity profiles for the decay in intensity due to photobleaching of the fluoroscein solution. This is achieved by measuring the decay due to photobleaching over time in a by CO$_2$ unaffected segment of the cylinder and correcting the measured intensities correspondingly. Finally, we once more normalise the obtained intensities by the maximum and minimum obtained intensities of the experiment which are found to be quite close for all experiments, resulting in $\overline I^*(z, \ t)$. 

\begin{figure}
	\centering
	\includegraphics[width=1.0\textwidth]{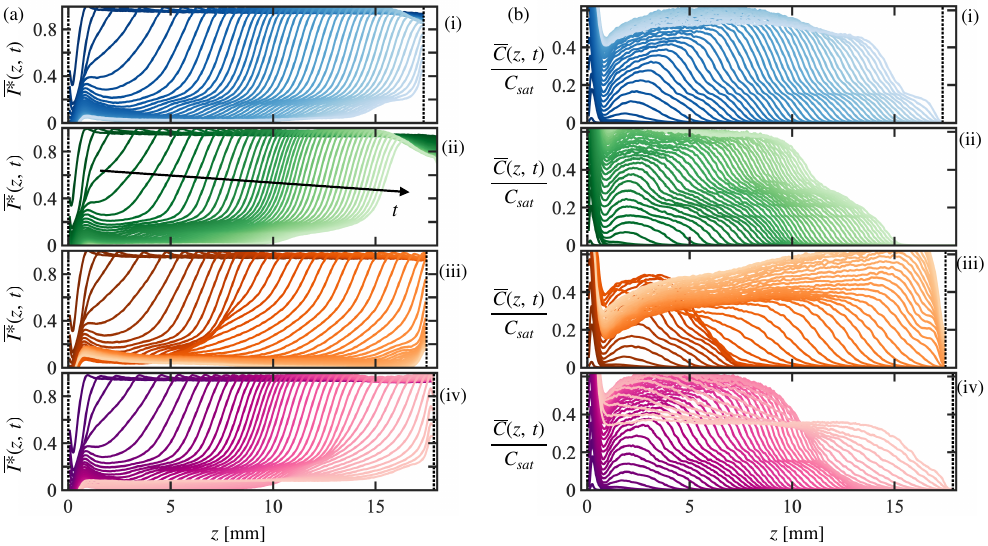}
	\caption{
	(a) Normalised intensity profiles for experiments (i--iv) from figure \ref{fig_snapshots}  and figure \ref{fig_front2d}.  The intensity profiles have been been horizontally-averaged over the entire cell diameter ($-d/2<x<d/2$) at every depth $z$. The time step between consecutive profiles is $ \Delta t = 20$ s for $0<t<15$ min.
	(b) Tentative concentration profiles directly obtained from $\overline I^*$ after calibration;  $\Csat$ refers to the saturation concentration. Concentration values close to the boundaries of the liquid column  are tainted by the presence of a meniscus (or a solid interface) and are limited to $\overline C /\Csat<0.6$,  which corresponds to the lower bound of the pH-sensitive range of sodium fluorescein. The vertical dotted lines indicate the depth of the top ($z = 0$) and bottom boundaries of the liquid barrier.
}
	\label{fig_profilesA}
\end{figure}

For experiments (i-iv), the obtained intensity profiles are shown in figure \ref{fig_profilesA}a. As before, the time step between consecutive profiles is $\Delta t = 20s$ and the profiles are shown for the entire experiment, i.e. between  $0 < t < 15$ min. The vertical dotted lines indicate the depth of the top ($z =0$) and bottom boundaries of the liquid barrier. For experiments (i) and (ii), the steady propagation discussed before is clearly reflected in the intensity profiles. As the front propagates through the barrier, the intensity drops rapidly, as expected based on the snapshots from figure \ref{fig_snapshots}. Moreover, the decrease in propagation velocity is again reflected in the intensity profiles, as the spacing between the profiles decreases as time progresses. For experiments (iii) and (iv), the shedding of the lateral buoyant upwelling plume causes the intensity at certain depth to increase, resulting in overlapping intensity profiles. 

As mentioned before, the emission intensity of fluorescein has a non-linear dependency on the pH level of the liquid. Therefore, we need to obtain a calibration curve before we can convert the measured intensities to the CO$_2$ concentration in the barrier. To achieve this, we performed a set of experiments in which the cylinder is placed in an inverted configuration. When inverted, the CO$_2$-liquid mixture is stably stratified and therefore the CO$_2$ can only be transported up the barrier by diffusion. We obtain the intensity profiles of these experiments and use these to obtain a calibration function $C/C_{sat} = F(1-I^*)$, linking the dimensionless concentration in the barrier to a measured intensity by means of the self-similar solution of the pure diffusion problem. A more detailed description of this process can be found in Appendix \ref{sec_appendixA}.

Figure \ref{fig_profilesA}b shows the resulting CO$_2$ concentration profiles for experiments (i-iv). Similarly, for experiments (v-xii), the intensity and concentration profiles can be found in the supporting material in figures S5 and S6. Note that we show the CO$_2$ concentration $\overline C(z ,\ t)$ as a fraction of the saturation concentration $C_{sat}$, with $\overline C /\Csat<0.6$,  which corresponds to the lower bound of the pH-sensitive range of sodium fluorescein. As a result, we cannot differentiate concentration levels $\overline C /\Csat>0.6$. The concentration values close to the boundaries of the liquid column are tainted by the presence of a meniscus (or a solid interface).

As expected, the concentration profiles of experiments (i) and (ii) show a steady progression of the CO$_2$ concentration in the liquid barrier. In experiments (iii) and (iv), the shedding of the lateral buoyant upwelling plume causes additional CO$_2$-rich liquid to be propagated downwards, while additional pure liquid is propagated upward by the plume. This is reflected by the concentration profiles, as we observe a sudden increase in concentration near the bottom of the cylinder, while the concentration decreases sharply at the top. Furthermore, an increase in mass transfer can also be observed after the shedding event, as the spatial spacing between the profile increases after shedding the buoyant plume. 

\begin{figure}
	\centering
	\includegraphics[width=0.5\textwidth]{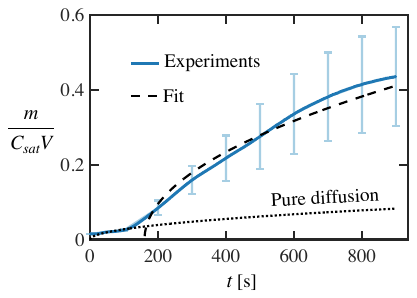}
	\caption{Mass $m$ of CO$_2$  absorbed by the liquid column as a function of time, averaged across all 12 experiments; $m$ is normalised by the maximum dissolution capacity $ \Csat V$, where $V \equiv \pi d^2 H/4$ is the liquid volume and $\Csat$  the saturation density (or concentration)  of CO$_2$ in water.
	The dissolution fraction is estimated from the 1D axial concentration profiles as  $m(t)/(\Csat V) = H^{-1} \int_0^H \overline C(z, \ t)/\Csat \: \dd z$.   The wide error bars reflect the variability between experiments and the uncertainties in concentration calibration.
The dissolution curve initially follows the self-similar solution for pure diffusion (dotted line): $m= 2C_\mathit{sat}A \sqrt{D t/\pi}$, where $A = \pi d^2/4$. At the onset of convection it starts to deviate from the purely diffusive behaviour due to the contribution of convection. In that regime the curve can be described by an effective diffusive behaviour $m= 2\Csat A \sqrt{D_c (t-t_c)/\pi}$ (dashed line), with the fit $D_c = 30 D$ as the effective diffusion coefficient and $t_c = 160$ s as the virtual time origin.
	}
	\label{fig_mass}
\end{figure}

Finally, we use the obtained CO$_2$ concentration profiles to calculate the total mass $m$ of CO$_2$ absorbed by the liquid barrier as a fraction of the maximum dissolution capacity. We obtain the dissolution fraction from the CO$_2$ concentration profiles as:

\begin{equation}
\frac{m(t)}{\Csat V} = H^{-1} \int_0^H \overline C(z, \ t)/\Csat \: \dd z,
\end{equation}

\noindent where $m(t)$ is the total mass of CO$_2$ in the liquid phase, $C_{sat}$ the saturation concentration of CO$_2$ in water, and $V \equiv \pi d^2 H/4$ the volume of the liquid barrier. As mentioned before, we can only measure the concentration up to $\overline C /\Csat<0.6$, as higher concentrations are outside the pH sensitive range of the fluorescein. Figure \ref{fig_mass} shows the obtained dissolution fraction  $m(t)/(\Csat V)$ versus time, averaged over all 12 experiments. The wide error bars reflect the variability among experiments, such as the occurrence of the shedding of a lateral buoyant upwelling plume, and the uncertainties in the concentration calibration. Initially, the dissolution curve follows the self-similar solution for pure diffusion (dotted line), where  $m= 2C_\mathit{sat}A \sqrt{D t/\pi}$ and $A = \pi d^2/4$. However, after the onset of convection, the curve deviates, resulting in $m= 2\Csat A \sqrt{D_c (t-t_c)/\pi}$ (dashed line) to be the best fit, with an effective diffusion coefficient of $D_c = 30 D$ and $t_c = 160$ s as the virtual time origin. Compared to other authors, who report finding $D_{c}/D \sim 10^2$ or $10 ^3$ for experiments conducted in varying PVT or Hele-Shaw cells, this seems reasonable as variations in experimental conditions and cell configuration differences appear to severely affect the obtained effective diffusion coefficients \cite{Du2019,Karimaie2017,Farajzadeh2009,Yang2019,Zhao2018}.

\section{Front propagation dynamics}
\label{sec:Frontpropagation}

\begin{figure}
	\centering
	\includegraphics[width=0.65\textwidth]{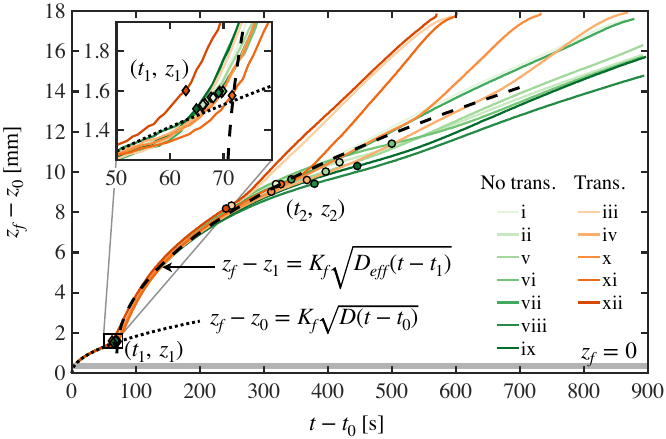}
	\caption{Front trajectory $z_f(t)$  corresponding to the 60 \%-intensity threshold of the horizontally-averaged intensity profiles (cf. figure \ref{fig_profilesA}), namely, $\overline I^*(z_f, \ t) = 0.6$. The front trajectory of all 12 experiments 	is plotted, which are colour-coded based on the absence (greens) or the occurrence (reds) of the shedding of the upwelling plume. The time $t$ and front position $z_f$ [the latter defined as the distance to the apex of the top meniscus] have been offset by the (fitted) virtual origin of the diffusive regime ($t_0 = 5.5 \pm 2.2$ s  and $z_0 = -0.3 \pm 0.06$ mm), which absorb the influence of the  finite curvature of the top meniscus and the typical flushing response time required for full exposure to the CO$_2$ ambient. Initially, $z_f(t)$ follows the self-similar solution for pure diffusion (dotted line), taking $K_f \equiv 2 \mathrm{erfc}^{-1}(C_f/\Csat) = 4.27$ corresponding to $C_f/\Csat = 2.5\times10^{-3}$ ($I^*= 0.6$). The onset of the convective instability (diamond markers) occurs at $t = t_1$, $z_1 = z_f(t_1)$, when the front acceleration ($\dd^2 z_f/ \dd t^2$) is maximum. A zoom-in is provided in the inset, which highlights the reproducibility of the time onset: $t_1-t_0 = 67.6 \pm 2.4$ s. Thereafter, $z_f(t)$ evolves in an enhanced diffusive manner  (dashed line), with an effective fitted diffusivity $D_\mathit{eff} = 8.25D$, which implies that  the convective velocity of the front decays in time. At approximately $t=t_2$ (circular markers), when $\dd^2 z_f/ \dd t^2 = 0$ for the first time, the front velocity stabilises and the front propagates as expected for late stage enhanced diffusive behaviour. For experiments in which a upwelling plume is shed, $t=t_2$ marks this moment and the velocity shoots off towards a higher velocity.}
	\label{fig_frontmaster}
\end{figure}

We carry on our analysis by focusing on the propagation dynamics of the CO$_2$ front in the liquid barrier. We define the position of the CO$_2$ front, $z_f(t)$, which we arbitrarily set to the 60$\%$-intensity threshold of the horizontally averaged intensity profiles, $\overline I^*(z_f, \ t) = 0.6$ (cf. figure \ref{fig_profilesA}). Tracing this position in time yields figure \ref{fig_frontmaster}, which shows the front trajectories of all 12 experiments versus time. The trajectories are offset by the (fitted) virtual origin $z_0 = -0.3 \pm 0.06$ mm and $t_0 = 5.5 \pm 2.2$ s of the diffusive regime. Correcting for the virtual origin absorbs the influence of the finite curvature of the top meniscus and the typical flushing response time required for full exposure to the CO$_2$. Furthermore, the front trajectories in figure \ref{fig_frontmaster} have been colour coded based on the absence (green) or the occurrence (red) of the shedding of an upwelling plume.

As explained before, the dissolution of CO$_2$ into the liquid barrier results in the formation of a boundary layer at the top interface. The mass transport in this layer is driven purely by diffusion and it is therefore unsurprising that the propagation of the front $z_f(t)$ follows the self-similar solution for pure diffusion:

\begin{equation}
\frac{ C(z, \ t)}{\Csat} = \mathrm{erfc}\left(\frac{z-z_0}{\sqrt{4D(t-t_0)}}\right).
\end{equation}
The front trajectory associated to concentration $C_f$ is thus
\begin{equation}
\label{eq_frontdiffusive}
z_f-z_0= K_f \sqrt{D(t-t_0)},
\end{equation}
where  growth prefactor $K_f$ depends on the concentration
\begin{equation}
\label{eq_Kfdiff}
K_f \equiv 2 \mathrm{erfc}^{-1}(C_f/\Csat).
\end{equation}

\begin{figure}
	\centering
	\includegraphics[width=0.75\textwidth]{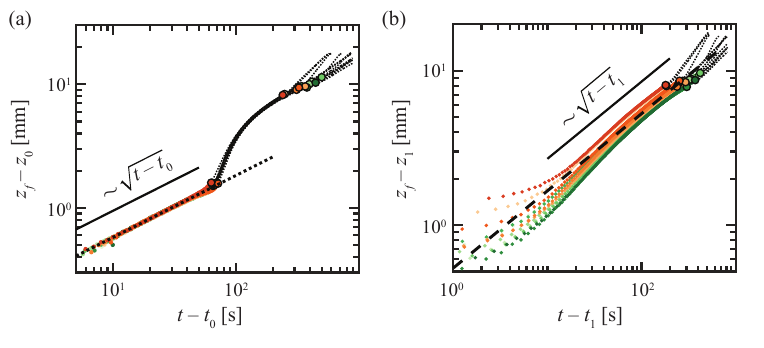}
	\caption{Double logarithmic plot of the two distinct regimes observed for the 12 front trajectories plotted in figure \ref{fig_frontmaster}: (a) the purely diffusive regime for $t<t_1$, and (b) the enhanced diffusive regime for $t_1<t$. 
	In (b), the convective virtual origin $(t_1, \ z_1)$ is off-scale and has been fitted accordingly, given that the front behaviour around $t_1$ is transitioning  (during a transition period of $\sim$10 s) and consequently does not offer a suitable virtual origin of the convective-decay regime. 
}
	\label{fig_frontregimes}
\end{figure}

\noindent  In figure \ref{fig_frontmaster}, the dotted line follows from Eqs. \ref{eq_frontdiffusive} and \ref{eq_Kfdiff}, taking $K_f \equiv 2 \mathrm{erfc}^{-1}(C_f/\Csat) = 4.27$ corresponding to $C_f/\Csat = 2.5\times10^{-3}$ ($I^*= 0.6$). We see that $z_f(t)$ follows the self-similar solution up to time $t_1$. We therefore define this first regime between $t_0 < t <t_1$ as the purely  diffusive regime. In figure \ref{fig_frontregimes}a, we show a rescaled plot of the purely diffusive regime on a double logarithmic scale, magnifying the $z_f(t) \sim \sqrt{t-t_0}$ scaling relation.

At time $t_1$, we observe a sharp acceleration of the CO$_2$ front position, due to the onset of the convective instability. Therefore, we can find t$_1$ and correspondingly $z_f(t_1) = z_1$ by finding the point in time at which the front acceleration, d$^2z_f/$d$t^2$, is maximum, indicated in figure \ref{fig_frontmaster} with the diamond markers. The inset shows a zoom-in around t$_1$ to show the reproducibility of the onset time $t_1-t_0 = 67.6 \pm 2.4$ s, and corresponding front depth $z_1 = 1.26 \pm 0.08$ mm.

At the onset of convection, the Rayleigh number based on the thickness $\delta(t)$ of the boundary layer is given by

\begin{equation}
\Ra_\delta(t)   \equiv \frac{\beta  \Csat g \delta^3(t)}{\nu D}. 
\label{eq:Ra_delta}
\end{equation}

\noindent  The height of the liquid barrier $H$ does not influence the onset. Taking the boundary layer thickness equal to the position of the front at the time we observe the onset of convection, i.e. $\delta = z_f(t_1) = z_1$, we obtain a critical value of $\Ra_{z_1} = (3.30 \pm0.6) \times 10^3 $, by taking the average critical value for the twelve experiments shown in figure \ref{fig_frontmaster}. We compare this value to the critical Rayleigh number from Ahlers {\it{et al.}} for Rayleigh-B\'enard convection in a cylinder with adiabatic sidewalls, which we believe to be the closest available approximation to our system \cite{Ahlers2022}:

\begin{equation}
\Ra_c   \equiv 1708 \left( 1+\frac{1.49}{\Gamma(t_1)^2}\right)^2,
\label{eq:Ra_c_ahlers}
\end{equation}

\noindent  where the local aspect ratio is defined as $\Gamma(t_1) =  d/z_f(t_1)$. However, we have to emphasise the differences between our system and the systems usually described in Rayleigh-B\'enard convection studies, for which (with constant $\Gamma$) equation \ref{eq:Ra_c_ahlers} holds. First of all, in those systems, it is assumed that at the onset of convection, a linear concentration (or temperature) profile exists as the base state which subsequently becomes unstable. Secondly, Rayleigh-B\'enard setups have reached a steady state (or are very close to such), while in our experiment the system has not and never will reach a steady state during our experimental time frame. Finally, a constant aspect ratio is assumed, while in our experiment the aspect ratio continuously decreases with time, since the front position $z_f(t)$ increases in time.

In our experiments, it is clear from figure \ref{fig_profilesA}b that we have non-linear concentration profiles in the boundary layer. As a result, the thickness of the self-similar diffusion boundary layer is not easy to define.  If we use $\delta$ to denote the effective thickness of the boundary layer, then $\delta  = K_\delta \sqrt{D(t-t_0)} $, where the growth prefactor $K_\delta \equiv 2 \mathrm{erfc}^{-1}(C_\delta/\Csat)$  depends on the choice of the concentration cut-off $C_\delta$, and hence $\delta$ may differ from the depth $z_f = K_f \sqrt{D(t-t_0)} $ of our chosen iso-concentration contour $C_f/\Csat= 2.5\times10^{-3}$. For example, when we calculate $Ra_c$ from equation \ref{eq:Ra_c_ahlers}, using $\Gamma(t_1) =  d/z_f(t_1) \approx 2.4$, we find $Ra_c = 2.71\times 10^3$, which is smaller than $\Ra_{z_1}$. The value of $\delta$ that satisfies $Ra_{\delta} = Ra_c$, i.e. for which equations \ref{eq:Ra_delta} and \ref{eq:Ra_c_ahlers} intersect, is exactly $\delta = \delta^* = 1.13$ mm, which is reasonably close to $z_1$, with a corresponding $Ra_{\delta^*} = Ra_c = 2.51\times 10^3$. This further emphasises the difficulty in defining the thickness for the self-similar diffusion boundary layer, as by selecting a lower intensity threshold, and thus higher concentration cut-off $C_\delta$, we could have reproduced the prediction from Ahlers {\it{et al.}} \cite{Ahlers2022}.

In addition, we compared our findings with the work of Tan and Thorpe, who also studied the dissolution of CO$_2$ in water \cite{Tan1992,Tan1999} and transient heat conduction in deep fluids \cite{Tan1996}. In their works, they try to account for the non-linear profile in the boundary layer within a theoretical framework which is compared with the experimental data. Using a PVT cell, they report an onset time of $t=100$ s for CO$_2$ dissolution in water. Deriving an expression for the maximum transient Rayleigh number, $Ra_{max}$, and taking $Ra_c = 1100$ (which holds for Rayleigh-B\'enard setups with a linear profile and upper free-surface \cite{Sparrow1964}), yielded the transition times with close agreement with experiments \cite{Tan1992}. When we enter the onset time obtained from our experiments, we find $Ra_{max} = 336$. This $Ra_{max}$ is much lower than $Ra_c$ = 1100, even though their setup is significantly wider than ours (d = 90 mm compared to our d = 3mm), which suggests that our critical Rayleigh number should be even higher than 1100. 

We conclude that applying the method from \cite{Tan1992,Tan1999} to obtain the onset time leads to a severe and unrealistic overestimation for our experiments, which could be connected to the fact that in \cite{Tan1992,Tan1999} the pressure response of the system was studied instead of directly comparing to the concentration profile. The plot reporting the onset time in Ref. \cite{Tan1992} comes with significant uncertainty. We note that adopting $t=64$s instead of $t=100$s, which still seems consistent with the data in Ref. \cite{Tan1992}, would result in similar findings to ours. Ref. \cite{Tan1992}, however, does indicate that it is quite difficult to define a precise critical Rayleigh number for experiments with a non-linear profile, as we also discussed above. We therefore conclude that our method of finding the critical critical Rayleigh number using equation \ref{eq:Ra_delta} and the boundary layer thickness at the onset gives the best approximation of the critical Rayleigh number for the chosen iso-concentration contour, which in our case is $C_f/\Csat= 2.5\times10^{-3}$.

After the onset of convection, the front propagates seemingly in a diffusive manner, however with an increased effective diffusion coefficient. By fitting all 12 experiments (dashed line in figure \ref{fig_frontmaster}) we find that $z_f(t)$ evolves with an effective fitted diffusivity $D_{eff} = 8.25D$. For comparison, Karimaie and Lindeberg report $D_{eff} = 5.8D$ for transport of CO$_2$ in water confined in porous media \cite{Karimaie2017}. Moreover, figure \ref{fig_frontregimes}b shows the same regime on a double logarithmic scale, highlighting the  $z_f(t)-z_1 \sim \sqrt{t-t_1}$ scaling relation. As a result, we define this regime as the enhanced diffusive regime. 

Eventually, the system appears to stabilise, leading to the front propagating at a seemingly stable terminal velocity. This moment, defined as the first time at which $\dd^2 z_f/ \dd t^2 = 0$, is referred to as $t = t_2$, marked with circular markers in figure \ref{fig_frontmaster}(b). While one could see this as a separate regime, it is in fact the late stage behaviour of the second regime. A more detailed explanation will be given in the next section. For experiments in which the shedding of an upwelling plume occurs, $t = t_2$ also happens to mark the moment at which this shoot off occurs. As mentioned before, if this event occurs, the front accelerates and propagates with a higher velocity in comparison to the experiments in which axisymmetry is not broken. As a result, curves with a second shedding event shoot off in figure \ref{fig_frontmaster} (red curves), although their scaling behaviour remains consistent with the experiments in which the shedding of the upwelling plume does not occur.

\section{Numerical model}
\label{sec:NumericalModel}

\subsection{Setup and governing equations}

\begin{figure}
	\centering
	\includegraphics[width=1.\textwidth]{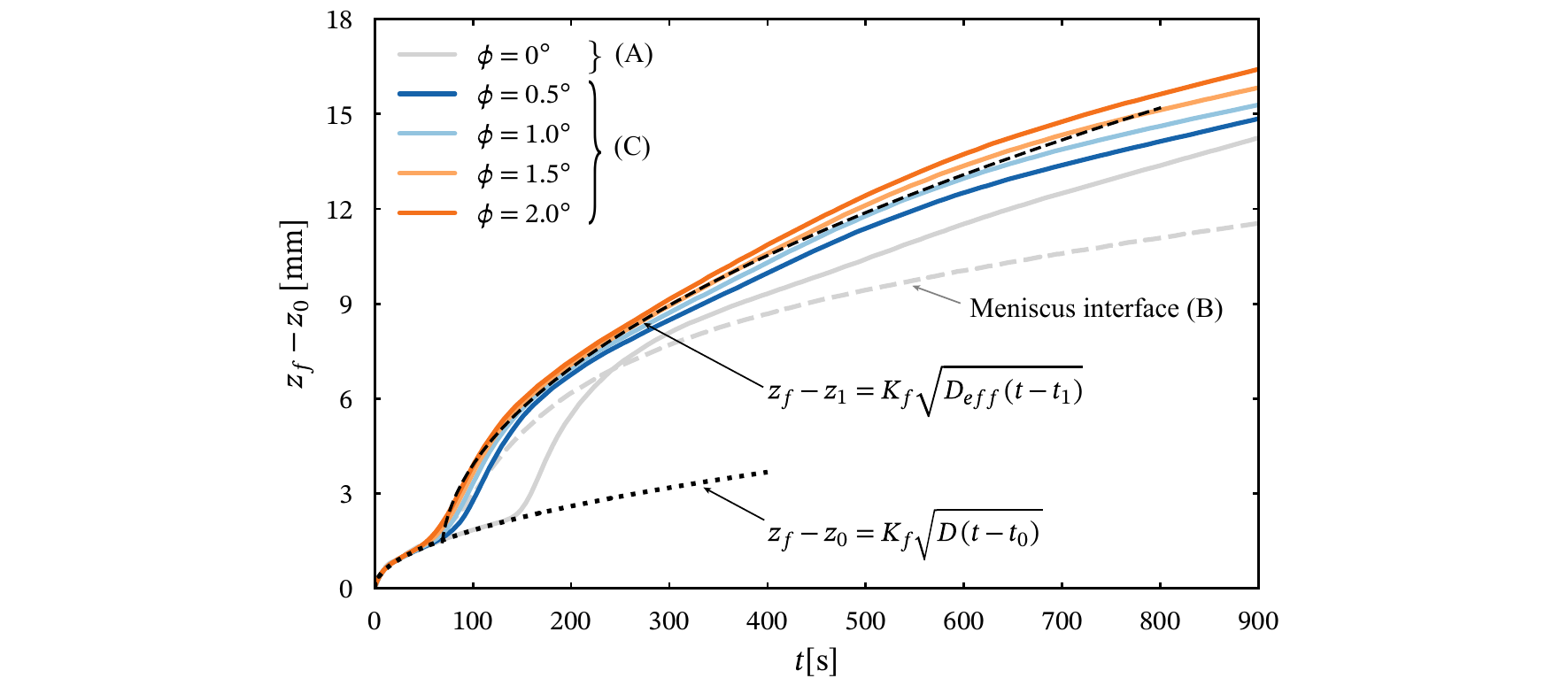}
	\caption{Front trajectory $z_f(t)$ corresponding to $\overline{C}(z)/C_{sat}= 2.5\times10^{-3}$ obtained via numerical simulations performed under different conditions, namely cases (A)-(C) as explained in the main text. The analytical solution of the purely diffusive problem (dotted black line) and the experimentally averaged trajectory after the onset of convection (dashed black line) have been provided for comparison.} 
	\label{fig_sims_front_time}
\end{figure}

\begin{figure}
	\centering
	\includegraphics[width=1.02\textwidth]{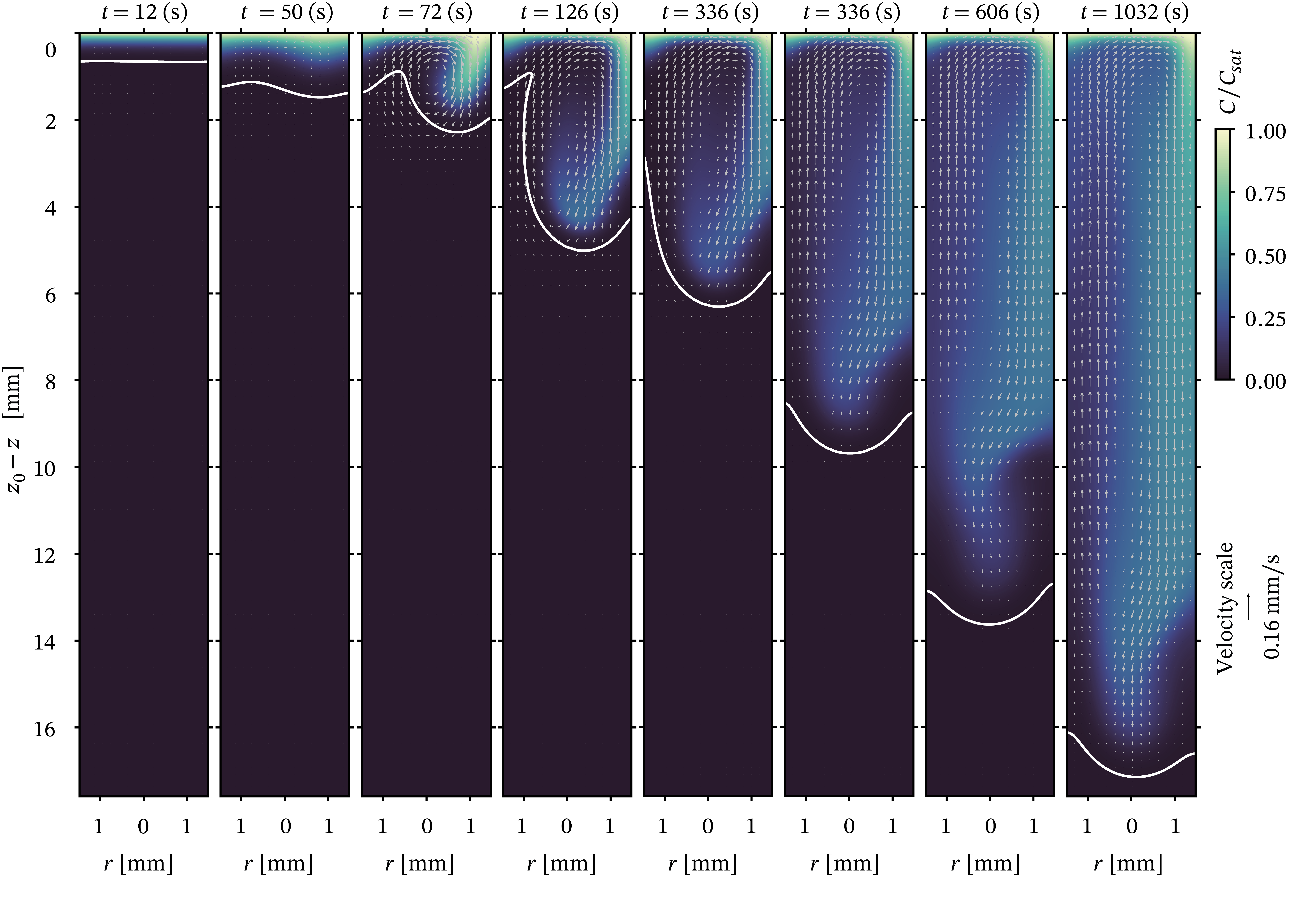}
	\caption{Time evolution of carbon dioxide concentration obtained by numerical simulations for case (C) with inclination angle of $\phi=1.5^\circ$. The setup has been tilted in the $\theta=0^\circ$ plane from which the snapshots have been taken. The white contour-lines show the front profile associated with $C_f/C_{sat}= 2.5\times10^{-3}$. Vectors denote the velocity field, the scaling of which has been provided in the figure.}
	\label{fig_sims_contourf}
\end{figure}

\begin{figure}
	\centering
	\includegraphics[width=0.3528\textwidth]{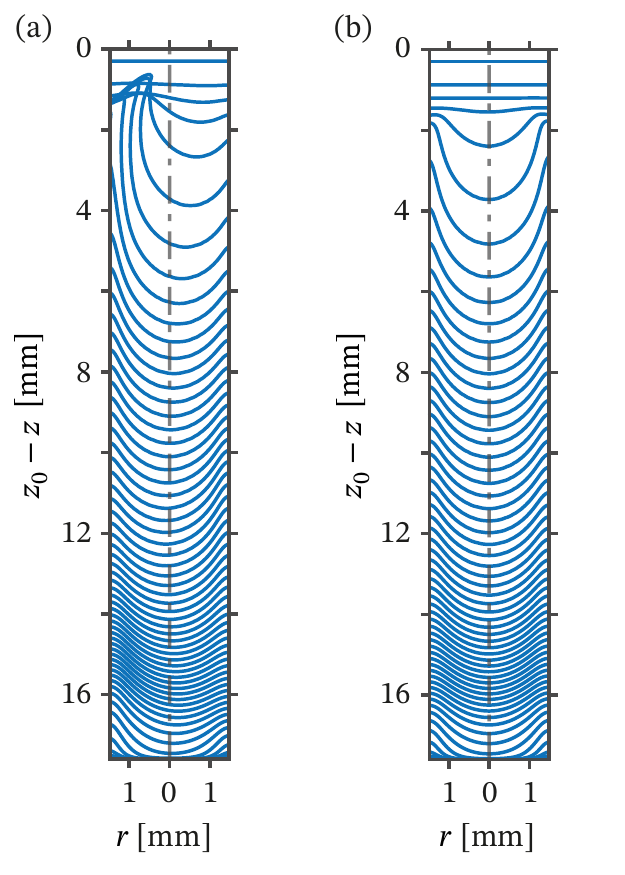}
	\caption{Propagation of the front isosurface obtained from numerical simulations for case (C) with inclination angle of $\phi=1.5^\circ$. The front isosurfaces correspond to $C_f/C_{sat}= 2.5\times10^{-3}$ and have been taken at (a) $\theta=0$, i.e., the inclination plane of gravity and (b) $\theta=90^\circ$ plane. Note that the propagation appears to be axisymmetric in the latter.}
	\label{fig_sims_isocontours}
\end{figure}

We continue our analysis by employing Direct Numerical Simulations (DNS) to unravel the physics governing the plume dynamics observed in the experiments. The numerical set-up, shown in figure \ref{fig_schematic}b, is a single-phase buoyancy-driven flow confined in a cylinder with an adiabatic sidewall and free-shear surfaces at top and bottom. The system is subjected to constant saturation concentration of carbon dioxide at the upper plate and zero concentration at the bottom. The dimensionless form of the advection-diffusion equation coupled with the three dimensional Navier-Stokes equations are employed under the incomprehensibility condition and the Oberbeck-Boussinesq approximation (in order to account for buoyancy forces caused by the (small) density variations);

\begin{subequations}\label{eq:AllNumerics}
\begin{equation}
    \label{eq:AdvectionDiffusion}
    \frac{\partial \tilde C}{\partial \tilde t} + \mathbf{\tilde u} \cdot \boldsymbol \nabla \tilde C = \frac{1}{\sqrt{RaSc}} \boldsymbol \nabla^2 \tilde C,
\end{equation}

\begin{equation}
    \label{eq:NavierStokes}
    \frac{\partial \mathbf{\tilde u}}{\partial \tilde t} + \mathbf{\tilde u} \cdot \boldsymbol \nabla \mathbf{\tilde u} = -\boldsymbol \nabla \tilde P + \sqrt{\frac{Sc}{Ra}}\boldsymbol \nabla^2 \mathbf{\tilde u} + \tilde C \hat{\mathbf e}_{\mathbf g} \cdot \hat{\mathbf e}_k,
\end{equation}

\begin{equation}
    \label{eq:Continuity}
    \boldsymbol \nabla \cdot \mathbf{\tilde u} = 0.
\end{equation}
\end{subequations}

\noindent Here, $\tilde C$, $\mathbf{\tilde u}$ and $\tilde P$ denote the dimensionless concentration, velocity, and pressure respectively. The height of the cylinder $H$, the carbon dioxide saturation concentration $C_s$, and the free fall velocity $\sqrt{g\beta C_s H}$ have been used for normalisation of the equations, where $g$ is the gravitational acceleration and $\beta$ the (isobaric and isothermal) volumetric concentration expansion coefficient. $\boldsymbol \nabla$ is the gradient operator in cylindrical coordinates, $\hat{\mathbf e}_{\mathbf g}$ is the unit normal vector in direction of the gravitational acceleration and $\hat{\mathbf e}_k$($k=z,r,\theta$) are the unit normal vectors pointing toward the axial, radial or azimuthal directions as shown in figure \ref{fig_schematic}c. The control parameters of the numerical model are the Rayleigh number $Ra_H$ and the Schmidt number $Sc$ as defined in section \ref{sec:Experimentalobservations}.  

The governing equations \ref{eq:AllNumerics} have been solved using a second-order accurate finite-difference scheme on a staggered grid and a fractional-step time-marching approach, the detail of which can be found in \cite{Verzicco1996}. Introducing a disturbance to the system is necessary to trigger the instabilities arising from the buoyancy driven convection. Hence, three different sources of disturbance are tested in the numerical simulations in order to find the most appropriate set-up which reasonably replicates the experimental observation. These are:

\begin{enumerate}[(A)] 
\item a perturbed initial concentration field with a random positive noise throughout the system, whose amplitude varies between zero and 1$\%$ of the carbon dioxide saturation concentration;
\item a meniscus liquid-gas interface at the top rather than a flat interfacial boundary. The meniscus shape was approximated by a cosine profile with a maximum depth of $0.01H$ at the centre of the domain. Saturation concentration and no-slip velocity conditions are enforced at the interface using an immersed boundary method based on linear interpolations as developed in \cite{Fadlun2000}; 
\item a slight tilt of the container relative to the direction of gravity, see figure \ref{fig_schematic}c. The tilting has been performed by rotating the gravitational acceleration vector $\mathbf{g}$ by the angle of $\phi$ with respect to the negative axial direction in $\theta=0$ plane. The inclination angles of $\phi=0.5^\circ, 1^\circ, 1.5^\circ~ \text{and}~ 2^\circ$ have been tested in the numerical simulations.
\end{enumerate}

Simulations have been conducted for the aforementioned cases (A)-(C) with different sources of disturbance. The grid resolution of $32\times 192 \times 256$ in radial, azimuthal, and axial directions respectively, similar to that of confined-rotating Rayleigh-Bénard convection \cite{Hartmann2022}, have been used after a grid independence check has been performed. Time marching has been achieved with variable times steps with a maximum of $d\tilde t = 2 \times 10^{-3}$ and $\text{CFL}=5 \times 10^{-1}$. The input of the simulations are the aspect ratio of the setup $\Gamma=d/H$, the height-based Rayleigh number $Ra_H$, and the Schmidt number $Sc$ which have been chosen as 0.1704, $8.8\times 10^{6}$, and 515, in accordance with the experiments.
some text:\message{width:\the\columnwidth} width:\the\columnwidth
\subsection{Numerical results}

The vertical location of the front corresponding to $\overline{C}(z)/C_{sat}= 2.5 \times 10^{-3}$, consistent with the analysis of the experimental results, has been plotted as a function of time in figure \ref{fig_sims_front_time} and compared to the experiments. The front location follows that of the pure diffusion problem in all cases up to the moment when convection sets in. Looking into the transition time, a remarkable discrepancy exists between the experiments and case (A) where the initial concentration field is perturbed. On the other hand, the agreement is reasonable for cases (B) and (C). For case (B), in which the top interface is modelled as a meniscus, it can be seen that, despite an accurate prediction of the onset time, the front velocity after the onset of convection is underestimated. Only the simulations from case (C), in which the setup is tilted and the shedding of the plume is therefore asymmetric, can reproduce the transient front location obtained in the experiments reasonably well. On this basis, we conclude that the dynamics are very sensitive to small tilt angles and that likely such a small misalignment also exists in the present experiments. Accordingly, we continue by analysing the results obtained by numerical modelling for case (C), specifically with an  inclination angle of $\phi=1.5^\circ$, which shows the best agreement with the experiments. For reference, the carbon dioxide concentration profile for case (B) with the meniscus interface can be found in the supporting materials. 

\begin{figure}
	\centering
	\includegraphics[width=1\textwidth]{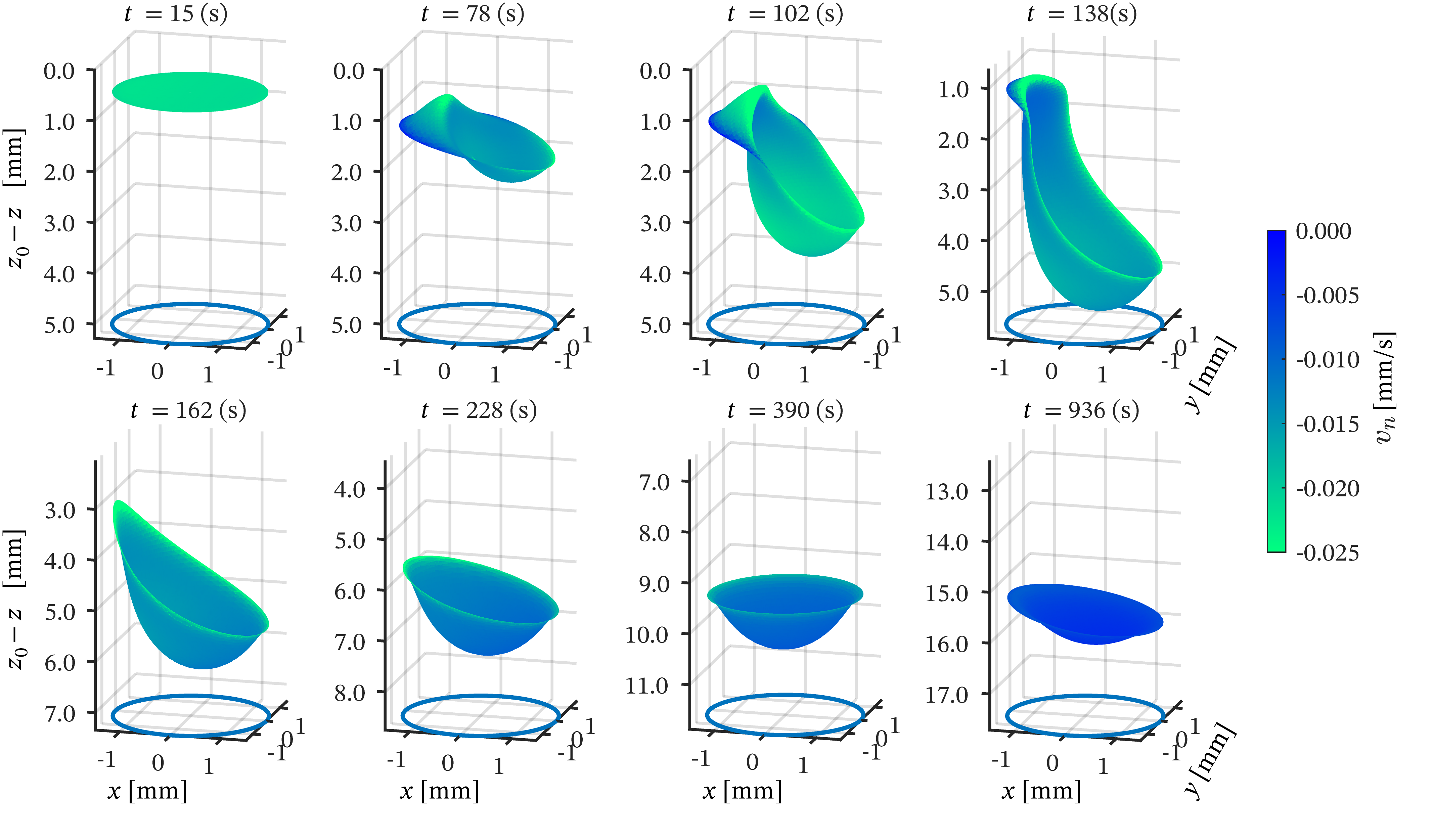}
	\caption{3D plot of the front isosurface evolving in time for case (C) with inclination angle of $\phi=1.5^\circ$. The isosurfaces correspond to $C_f/C_{sat}= 2.5\times10^{-3}$. The colourmap indicates the local propagation velocity of the front $v_n$ normal to the isosurface calculated according to equation \ref{eq_Vn1}. }
	\label{fig_sims_isosurfaces}
\end{figure}

Figure \ref{fig_sims_contourf} shows the simulation snapshots for the carbon dioxide concentration profile superimposed with the front isocontour corresponding to $C_f/C_{sat}= 2.5 \times 10^{-3}$ and vectors representing the velocity field. The shown slices correspond to the inclination plane $\theta=0$, the plane in which also the gravity vector is tilted. Initially, the front isocontour propagates as a horizontal line, following the analytical solution of the pure diffusion problem as shown in figure \ref{fig_sims_front_time}. Around $t \sim$ 1 minute, as observed in figure \ref{fig_sims_front_time}, convection sets in and as a result a vortex forms above the front whose direction is clockwise and consistent with the tilting direction of the setup. Therefore, the front shape gets distorted complying with the flow structure forming behind. The generated convective flow remains active behind the front during its entire evolution from top to the lower boundary of the setup and its deformed shape at very low concentrations, suggests a complex concentration field in the solution. Remarkable asymmetry in the front profile, particularly at the transition time, is observed from the simulations result as opposed to the experimental measurements. However, this also strongly depends on the angle of the view. The shape evolution of the front isocontour, shown in figure \ref{fig_sims_isocontours} for two different 2-dimensional slices corresponding to $\theta=0^\circ$ and $\theta=90^\circ$, indicates axisymmetric or asymmetric profiles depending on the frame of reference chosen. The 3D shape of the front isosurface has also been plotted in time in figure \ref{fig_sims_isosurfaces}, which will be discussed in more detail in the next section in order to clarify the physical mechanisms governing the interface dynamics.

\begin{figure}
	\centering
	\includegraphics[width=0.5\textwidth]{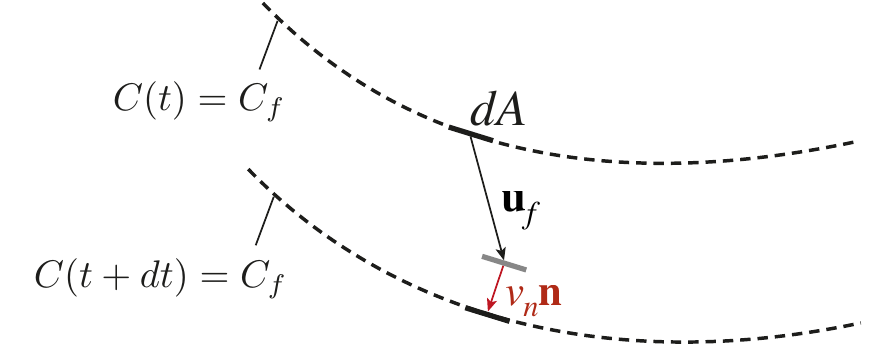}
	\caption{Graphical representation of the local propagation velocity of the front isosurface.}
	\label{fig_sketch_vn}
\end{figure}

\subsection{Front local propagation velocity}

Next, we consider the propagation of the front in more detail. As shown schematically in figure \ref{fig_sketch_vn}, the velocity ($\textbf{u}_{iso}$) of an interface element $dA$ of the front can be decomposed into a component due to the advection of the underlying fluid element and a propagation relative to the latter ($\textbf{V}$), such that  $\textbf{u}_{iso}=\textbf{u}_f + \textbf{V}$. By definition, the interface propagation is normal to the iso-surface such that 
$\textbf{V}=v_n \hat{\textbf{n}}$, with the surface unit normal vector $\hat{\textbf{n}}=\boldsymbol \nabla C / \mid \boldsymbol \nabla C \mid$.
Following an approach previously employed for enstrophy iso-surfaces in turbulent flows \cite{Holzner2011,Wolf2012}, iso-scalar surfaces in turbulent scalar mixing with chemical reactions \cite{Dopazo2006} and flame propagation in combustion problems  \cite{Chakraborty2007}, we can derive an expression for the interface propagation velocity $v_n$ by noting that in a frame of reference moving with the iso-surface element, the total rate of change of concentration is zero.
This leads to
\begin{equation}\label{eq_MatDertivative}
    \frac{D^s C}{D^s t}= \frac{\partial C}{\partial t} + \textbf{u}_{iso} \cdot \boldsymbol \nabla C =\frac{\partial C}{\partial t} + \left( v_n \hat{\textbf{n}} + \textbf{u}_f\right) \cdot \boldsymbol \nabla C = 0,
\end{equation}
which can be solved for $v_n$ to yield
\begin{equation}\label{eq_Vn1}
    v_n=-\frac{\frac{\partial C}{\partial t}+\textbf{u}_f \cdot \boldsymbol \nabla C}{\mid \boldsymbol \nabla C\mid} = -\frac{\frac{D C}{Dt}}{\mid \boldsymbol \nabla C \mid}
    =-\frac{D \boldsymbol \nabla^2 C}{\mid \boldsymbol \nabla C\mid}.
\end{equation}

Given the incompressibility of the fluid, advection does not affect the mean interface position. The front propagation is therefore solely related to $v_n$ and therefore diffusive in nature at all times. A quantitative relation can be obtained by equating the volume flux across the convoluted interface to that through the mean interface according to \cite{Krug2015,VanReeuwijk2014} 
 \begin{equation}\label{eq:Q}
     Q\equiv \int_{A_{iso}}v_n \dd A = A_0\frac{\dd z_m}{\dd t},
 \end{equation}
where the integration is over the surface area $A_{iso}$ of the iso-contour. Using an average of $v_n$ across $A_{iso}$ denoted by the overbar, this leads to 
\begin{equation}\label{eq_dzmdt_A_vn}
    \frac{\dd z_m}{dt}=\frac{A_{iso}}{A_0}\overline{v}_n, 
\end{equation}
which now expresses the mean front propagation as diffusive propagation ($\overline{v}_n$) amplified by interface convolutions. Note that here we use $z_m$ to denote the mean position of the isosurface instead of $z_f$ in order to distinguish the volume average implied by equation \ref{eq:Q} from the 2D average used for $z_f$.

\begin{figure}
	\centering
	\includegraphics[width=1\textwidth]{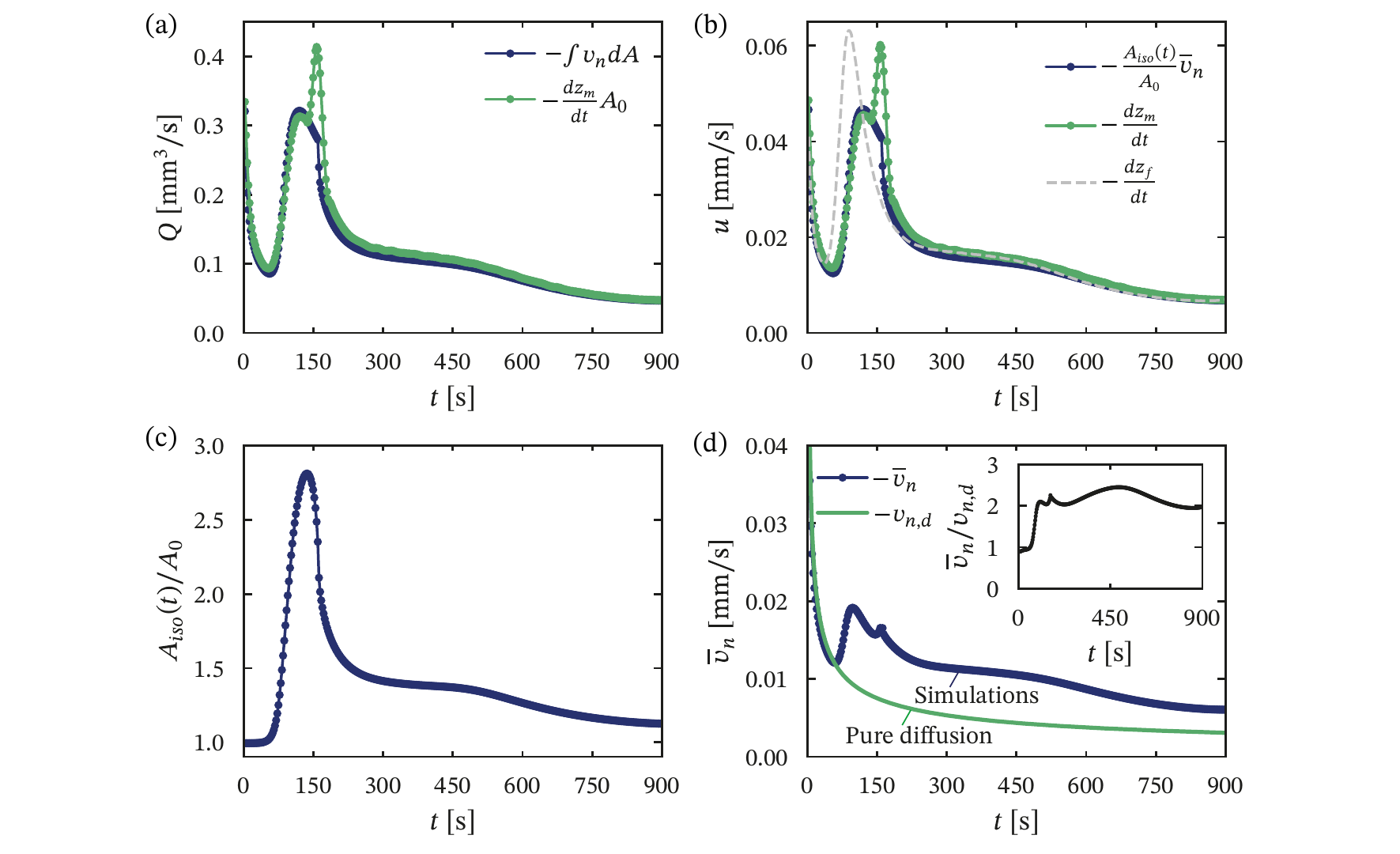}
	\caption{(a) Volumetric flux $Q$ of the local propagation velocity $v_n$ across the front iso-surface, compared to the rate of change in the mean volume above the iso-surface, $A_0 \dd z_m/\dd t$. (b) Front propagation velocity computed from the time evolution of the cross-sectional average of the 3D iso-surface, $dz_m/dt$, and from the product of the relative surface area and averaged local propagation velocity. $\dd z_f/\dd t$ is provided where $z_f$ is the front trajectory obtained from the horizontally-averaged concentration profile in 2D slices, in accordance with the front tracking approach in experiments. (c) Relative surface area of the iso-surface with respect to the cross sectional area of the cylinder, $A_0$. (d) Temporal evolution of the averaged local propagation velocity, $\overline{v}_n$, of the iso-surface. $\overline{v}_n$ for the iso-surface in pure-diffusion problem, compared to the pure diffusive case, $v_{n,d}$, where it equals the front propagation velocity $\dd z_m/\dd t= K_f \left[ D(t-t_0) \right]^{-1/2}$, with $K_f$ defined in equation \ref{eq_Kfdiff}. The ratio $\overline{v}_n/v_{n,d}$, is shown in the inset.}
	\label{fig_sims_v_stats}
\end{figure}

\noindent  
Figure \ref{fig_sims_isosurfaces} shows the front corresponding to the isosurface at $C/C_{sat} = 2.5 \times 10^{-3}$ at different moments in time and color-coded with the local magnitude of $v_n$. From these snapshots it becomes clear how convection significantly enhances the interfacial surface area, which is then decreased again as a result of the interface propagation. Convection is also seen to increase $v_n$ locally, in particular around $t = 100s$, which is due to a steepening of the scalar gradients close to the front (see also figure \ref{fig_sims_contourf}). Here it should be noted that $\mathbf{\hat n}=\boldsymbol \nabla C/\mid \boldsymbol \nabla C\mid$ always points toward the region with higher concentrations, i.e., the region above the front iso-surface, such that negative values of $v_n$ correspond to an outward propagation of the interface.

Results for computing the volume flux across the interface from integrating $v_n$ ($Q$) and based on evaluating $A_0dz_m/dt$ are compared in figure \ref{fig_sims_v_stats}(a). As can be seen the agreement is very good as expected, apart from a short period around $t \approx 150$ s. During this time, the wall-parallel part of the iso-surface reaches the wall (see figure \ref{fig_sims_isosurfaces}), leading to a sudden decrease in the front surface area as well as in its mean location, which is not captured sufficiently accurately by our method to extract the isosurface.

The panel in figure \ref{fig_sims_v_stats}b presents the same data as figure \ref{fig_sims_v_stats}a, only this time expressed as effective mean front velocities according to (\ref{eq_dzmdt_A_vn}). This form enables a direct comparison to the front velocity based on $z_f$,  previously shown in figure \ref{fig_sims_front_time}. For most times the agreement between $dz_f/dt$ and $dz_m/dt$ is good, but significant difference arise for the peak following the onset of convection. This highlights that especially during this period ($70\lessapprox t \lessapprox 200$) it is important to account for the three-dimensionality of the flow.

The volumetric fluxes and propagation velocity of the front are similarly calculated for the simulations in case (B), where the top boundary is modelled with the meniscus interface, and have been compared in figures S8a and b in the supporting materials. The agreement at the times when the iso-surface attains its maximum surface area is better for that case, due to less interaction of the iso-surface with the wall (see figure S7 in the supporting materials).

For a closer analysis we disentangle the effects leading to diffusion enhancement, followed by faster propagation of the front after transition to convection. Equation \ref{eq_dzmdt_A_vn} encompasses the two key parameters contributing to the front dynamics; the relative area of the iso-surface, $A_{iso}/A_0$, multiplied by the averaged local propagation velocity, $\overline{v}_n$, which together define the rate of diffusive transport across the front interface. Any changes in the front surface area or the concentration gradients (diffusive fluxes) in its vicinity can ultimately alter the diffusion rate across the front iso-surface and impact the propagation velocity. Therefore, we plot the time evolution of $A_{iso}/A_0$ and $\overline{v}_n$ in figures \ref{fig_sims_v_stats}c and d, respectively. The front velocity in the pure diffusion problem, which is solely equivalent to $\overline{v}_n$ (the front remains always flat and thus the relative surface area is unity in equation \ref{eq_dzmdt_A_vn}), has analytically been obtained from equation \ref{eq_frontdiffusive} as $v_{n,d}=K_f D (t-t_0)^{-1/2}$ and plotted in figure \ref{fig_sims_v_stats}d. The relative local propagation velocity of the front with respect to that of the pure diffusion problem, i.e., $\overline{v}_n/v_{n,d}$, is also shown in the inset.

Up to transition time, the front iso-surface remains flat, meaning that the relative surface area does not change in this period and thus remains equal to unity. Similarly, the local propagation velocity follows that of the pure diffusion problem as they are essentially the same before the onset of convection, which leads to a relative local propagation velocity equal to unity, as indicated in the inset. The value of $\overline{v}_n$ decreases within this period, complying to the front dynamics governed by diffusion regime, i.e. $dz_m/dt \sim t^{-1/2}$. 

Once convection sets in, the underlying flow field distorts the front intensely, as demonstrated in figures \ref{fig_sims_contourf} and \ref{fig_sims_isosurfaces} around $t \approx 1$ min. This leads to a significant increase in the surface area of the front, see figure \ref{fig_sims_v_stats}c, as well as the adjacent concentration gradients, $\boldsymbol \nabla C$. The latter is reflected in local propagation velocity when it increases after t $\approx$ 1 min in figure \ref{fig_sims_v_stats}d ($\overline{v}_n \sim \boldsymbol \nabla^2 C$ and $\boldsymbol \nabla^2 C$ is higher in the vicinity of a stretched interface since $\boldsymbol \nabla C$ has non-zero components in lateral directions). As a result, lateral diffusive fluxes across the interface intensify, which act against the further convolution of the front and lead to a re-flattening process, once a maximum surface area is reached. Consequently, the front surface area and local propagation velocity drop after the maximum distortion until a ``steady state" is reached (figures \ref{fig_sims_v_stats}c and d). They determine the trend of the total propagation velocity shown in figure \ref{fig_sims_v_stats}b. At this point, the advective fluxes causing the front distortion approximately equal the lateral diffusive fluxes across the interface which leads to much lower temporal variations afterwards.

The equilibrium state is even more evident in the simulations for case (B), with the meniscus interface at the top boundary. In these simulations, the distortion of the front is less pronounced and therefore the front interface has enough time to almost completely go through the re-flattening process, as shown in figure S7 at $t$=1737 s. Moreover, the relative front surface area and local propagation velocity, depicted in figures S8c and d, approach the values close to unity, meaning that the dynamics pertinent to the pure-diffusion regime are almost recovered and the front interface propagates with a nearly constant velocity at late times.

Therefore, although the emerged flow field after the onset of convection does not directly impact the mean location of the front, it does play a significant role in amplifying the carbon dioxide diffusion rate across the front interface through increasing the front interface surface area and local concentration gradients. The front iso-surface accelerates remarkably after the onset of convection and the front trajectory after the onset can still be described approximately with a relation similar to the analytical solution of a pure-diffusion problem as described in equation \ref{eq_frontdiffusive}, albeit with an effective diffusion constant $D_{\textrm{eff}}$, which accounts for the enhanced diffusion observed in the convective regime. For the case specifically studied here, the post-transition front trajectory, namely the enhanced diffusive regime, can be approximated with $D_{\textrm{eff}}=8.25D$, plotted for comparison with the data obtained via experiments and numerical simulations in figures \ref{fig_frontmaster} and \ref{fig_sims_front_time}, respectively.  

\section{Conclusions}
\label{sec:conclusions}

We have investigated the dissolution and subsequent propagation dynamics of carbon dioxide gas into a liquid barrier confined to a vertical glass cylinder, both experimentally and through direct numerical simulations. Replacing the ambient air above the cylinder with a CO$_2$ atmosphere, induces the dissolution of CO$_2$ into the liquid barrier. Initially, the dissolution of CO$_2$ results in the formation of a CO$_2$-rich water layer, which is denser in comparison to pure water, at the top gas-liquid interface. While initially stable, continued dissolution of CO$_2$ into the water barrier results in the layer becoming gravitationally unstable, leading to the onset of buoyancy driven convection and, consequently, the shedding of a buoyant plume. By adding sodium fluorescein, a pH-sensitive fluorophore, we directly visualise the dissolution and propagation of the CO$_2$ across the liquid barrier. Tracking the CO$_2$ front propagation in time allows us to define two clear propagation regimes.

At first, before the onset of convection, the growth dynamics of the boundary layer are purely governed by diffusion (the diffusive regime). The Rayleigh number continues to increase until it reaches the critical value of our system of $\Ra_{z_1} = (3.30 \pm0.6) \times 10^3 $ and convection starts. After the onset of convection, the propagation dynamics of the CO$_2$ front appear to also behave diffusively, albeit with an effective diffusion coefficient 8.5 times larger than expected for CO$_2$ in water. This enhanced diffusive regime remains throughout the experiments, until the system either reaches a ``steady state", at which point the front propagates at a constant velocity until it reaches the bottom interface, or becomes unstable, leading to the shedding of an upwelling plume and accelerating towards a higher velocity.

Using direct numerical simulations, we have uncovered the roots of the observed propagation mechanics. Initially, before the onset of convection, the simulations show that the relative surface area of the CO$_2$ front does not increase and the local propagation velocity follows the expected trend for a purely diffusive problem. After the onset of convection, first the relative surface area and local concentration gradients incorporated in the averaged local propagation velocity on the front, $\overline{v}_n$, concurrently increase due to the emerging local fluid flow. As a result the diffusive transport rate across the front interface is remarkably amplified, leading to much faster propagation velocity of the CO$_2$ front. In the meantime, increased lateral diffusive fluxes across the distorted interface act as a competing mechanism against the advective fluxes and further convolution of the front. This triggers the re-flattening process of the CO$_2$ front as a result of which the front surface area and local propagation velocity drop and a ``steady state” is reached. At this point, the advective effects causing the front distortion approximately equal the diffusive flattening of the interface, resulting in the front propagating at a seemingly constant velocity. The front trajectory after the onset can still be described with a relation similar to the analytical solution of a pure-diffusion problem, albeit with an effective diffusion 8.5 times higher than expected for CO$_2$ in water.
 
Our findings offer insight into the mass-transfer effects encountered in laterally confined CO$_2$ sequestration operations, as well as microfluidic or microreactor devices comprising segmented gas-liquid systems or density-changing solutes. Such a better understanding of the formation and propagation dynamics of the convective plume can uncover previously undiscovered mechanics pertaining to the dissolution and mixing off chemical species in a variety of applications. An interesting and relevant route to follow is the extension of our work to vessels with larger lateral extension (larger aspect ratio), where many plumes drive the downwards transport of the flow. Based on the results of Shishkina \cite{Shishkina2021}, we expect a strong increase of the transport with increasing aspect ratio. For very large aspect ratios, the unconfined limit of CO$_2$ sequestration will be approximated \cite{DePaoli2021}.

\section*{Acknowledgements}
We thank Robert Hartmann for assisting in the numerical simulations. This work was supported by the Netherlands Center for Multiscale Catalytic Energy Conversion (MCEC), an NWO Gravitation programme funded by the Ministry of Education, Culture and Science of the government of the Netherlands. This project has received funding from the European Research Council (ERC) under the European Union's Horizon 2020 research and innovation programme (grant agreement No. 950111, BU-PACT). This project has received funding from the European Union’s Horizon 2020 research and innovation programme under the Marie Skłodowska‐Curie (grant agreement No. 801359). We acknowledge PRACE for awarding us access to MareNostrum at Barcelona Supercomputing Center (BSC), Spain, and Joliot-Curie at GENCI@CEA, France (Project No. 2021250115). 

\appendix

\section{Intensity profile normalisation and concentration calibration}
\label{sec_appendixA}
The grayvalue (green channel value) profile $G(z, \ t)$ of the raw fluorescence images  is first normalized by the initial profile, i.e. without significant amounts of CO$_2$ present in the liquid, to correct for the spatial inhomogeneity of the LED lighting. Thus,
$I = G(z,\ t)/G(z, \ 0)$ is the normalized apparent intensity. However, $I$ decays exponentially in time due to photobleaching. We assume that rate of change of the $I$ is the sum of the rate of change due to CO$_2$-quenching (pH change) and the rate of change imposed by photobleaching:
\begin{equation}
\frac{\mathrm{d}I}{\mathrm{d}t} = 
\frac{\mathrm{d}\hat I}{\mathrm{d}t} 
-\beta \hat I,
\label{pb}
\end{equation}
where $\hat I$ denotes the true intensity (corrected for photobleaching) and $\beta$ is the photobleaching rate constant.
It follows that $\beta$ is in fact pH-dependent, i.e., $\beta = \beta(\hat I^*)$.
We approximate the dependence to be linear as $\beta = a\hat I + b$, where coefficients $a$ and $b$ are obtained experimentally, e.g. from the intensity decay rate within the CO$_2$-free  region in the water column in combination with measurements from purely diffusive experiments (where the cell is inverted).  
We find $a = -5.0 \pm 0.2 \times 10^{-4}$ s$^{-1}$ and $b = 6.8 \pm 0.3 \times 10^{-4}$ s$^{-1}$ across the 12 experiments.
The corrected intensity at any location $z$ can be solved  for iteratively by linearising  Eq. (\ref{pb}) as follows:
 \begin{equation}
\hat I(z)_{n+1}-\hat I(z)_{n} = I_{n+1}(z)- I_{n}(z) +\beta_n(z) \hat I_n(z)\Delta t,
\end{equation}
where subscript $n$ refers to the current time step or image frame and $\Delta t$ is the time difference between consecutive time steps.
Finally, the corrected intensity is then renormalised by the maximum and minimum intensity values (within the central region far from the shadowing effect of the meniscii or solid boundaries). Thus,
\begin{equation}
 I^*(z, \ t) = \frac{\hat I(z,\ t)-\mathrm{min}(\hat I)}{\mathrm{max}(\hat I)-\mathrm{min}(\hat I)},
\end{equation}
keeping in mind that $\mathrm{max}(\hat I) \approx 1.0$ and $\mathrm{min}(\hat I) \approx 0.4$ are quite close (ideally identical) for all experiments.

\begin{figure}
	\centering
	\includegraphics[width=0.4\textwidth]{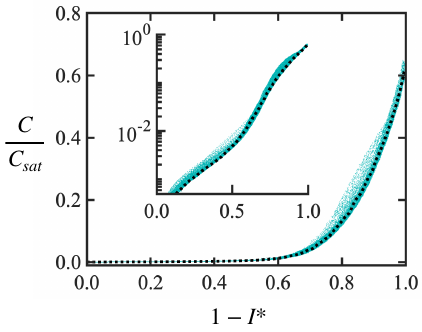}
	\caption{Calibration curve (dotted line), which relates  the normalised intensity $I^*$  to the dimensionless concentration, $C/\Csat$ under our particular experimental conditions.  The blue data points are measurements of the self-similar intensity profile $I^*(\eta)$ of the calibration (pure diffusion) experiment, for which $C(\eta)/\Csat = \mathrm{erfc}(\eta)$. Inset: same calibration curve plotted in semi-logarithmic axes.}
	\label{fig_cal}
\end{figure}

For the concentration calibration, we relate the intensity profile of a diffusive experiment to the self-similar profile $C/\Csat = \mathrm{erfc}(\eta)$, with $\eta = z/\sqrt{4Dt}$. Consequently, the intensity profiles evolves self-similarly too: $I^*(z, \ t)$ collapse into the same curve $I^*(\eta)$. Thus, the calibration function $C/\Csat = F(1-I^*)$ can be obtained by a monotonic fit  on a plot of $\mathrm{erfc}(\eta) $ vs. $1-I^*(\eta)$, which is provided in figure \ref{fig_cal}.

\clearpage

%

\end{document}


\title{Supplementary material for: Diffusive and convective dissolution of carbon dioxide in a vertical cylindrical cell}

\author{Dani\"{e}l P. Faasen}
\thanks{These two authors contributed equally.}
\email{d.p.faasen@utwente.nl}
\author{Farzan Sepahi}%
\thanks{These two authors contributed equally.}
\email{f.sepahi@utwente.nl}
\author{Dominik Krug}%
\email{d.j.krug@utwente.nl}
\author{Roberto Verzicco}%
\email{r.verzicco@utwente.nl}
\author{Pablo Pe\~{n}as}%
\author{Detlef Lohse}%
\email{d.lohse@utwente.nl}
\author{Devaraj van der Meer}%
\email{d.vandermeer@utwente.nl}

\affiliation{%
Physics of Fluids Group, Faculty of Science and Technology, University of Twente, P.O. Box 217, 7500 AE Enschede, The Netherlands\\
}%

\date{\today}
\maketitle

\begin{figure}
	\centering
	\includegraphics[width=0.85\textwidth]{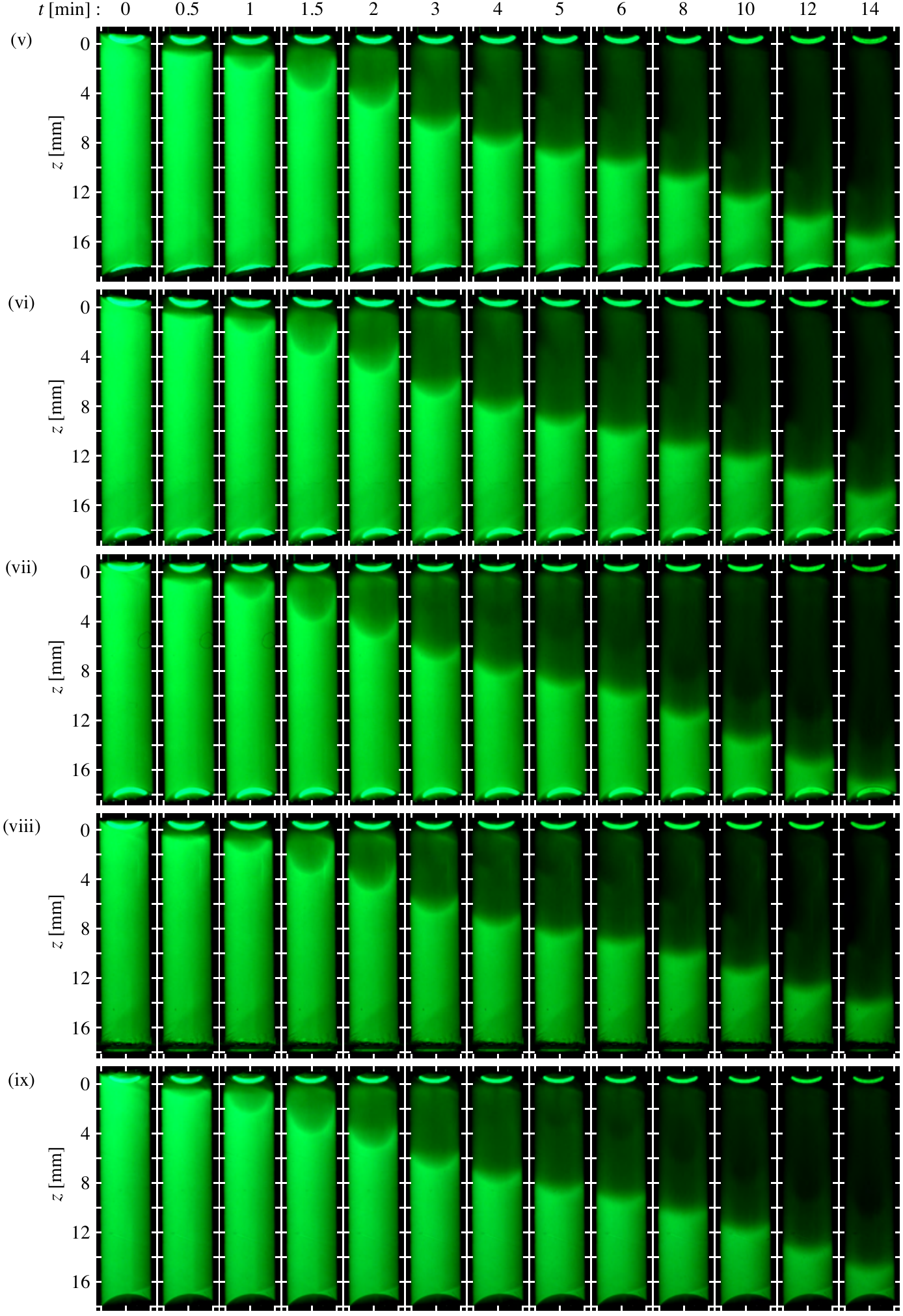}
	\caption{Snapshots of experiments (v--ix) where no mode transition is observed. Note that the bottom interface of experiments (v--vii) is liquid--gas, for (viii) it is liquid--solid, and for (ix) it is liquid--liquid (water--n-hexadecane).  See caption of figure 2 for  more details.}
	\label{fig_snapshotsb}
\end{figure}

\begin{figure}
	\centering
	\includegraphics[width=0.85\textwidth]{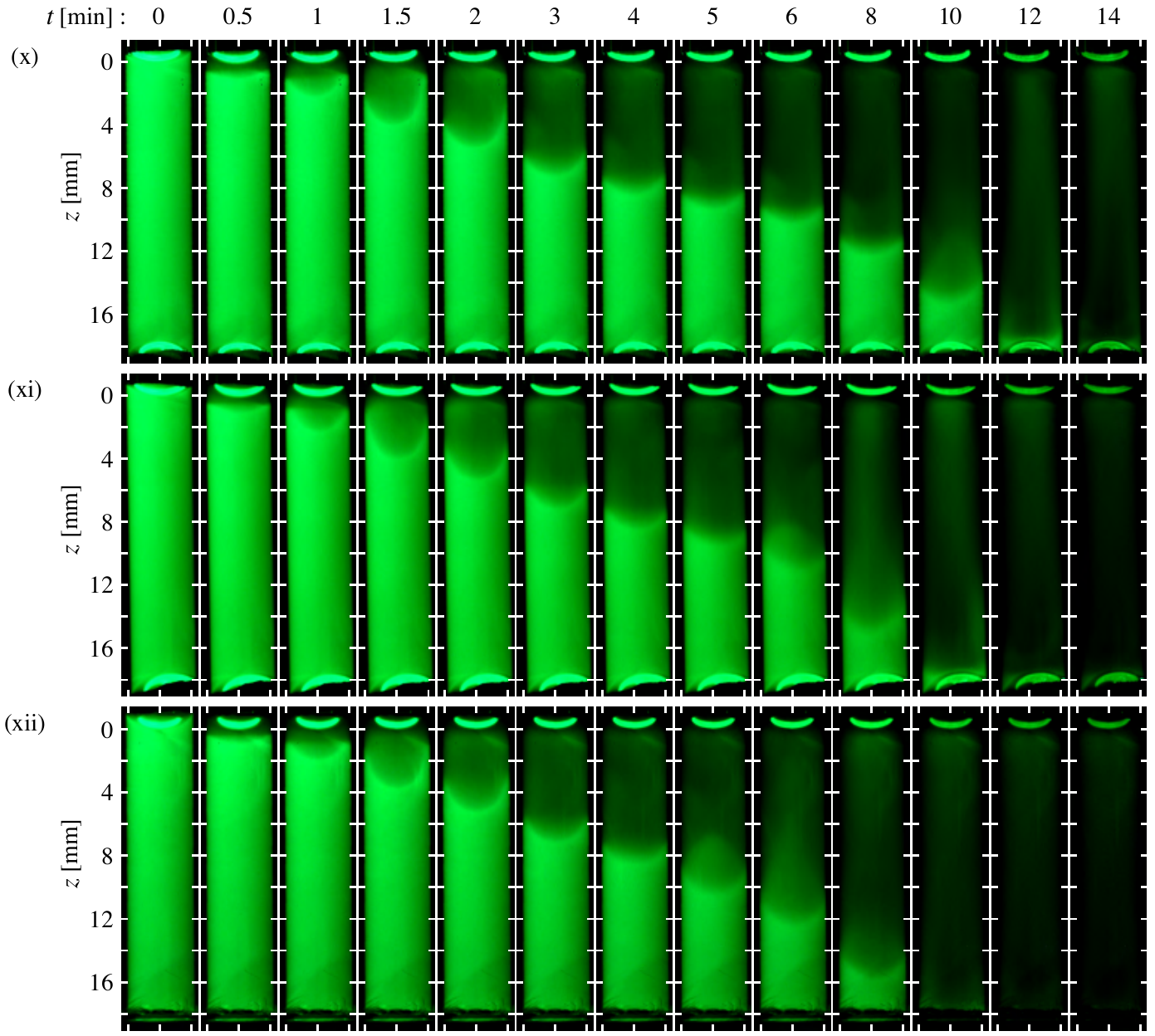}
	\caption{{Snapshots of experiments (x-xii) where mode transition is present. Note that the bottom interface of experiments (x, xi) is liquid--gas, for (xii) it is liquid--solid. See caption of figure 2 for  more details.}}
	\label{fig_snapshotsc}
\end{figure}

\begin{figure}
	\centering
	\includegraphics[width=0.7\textwidth]{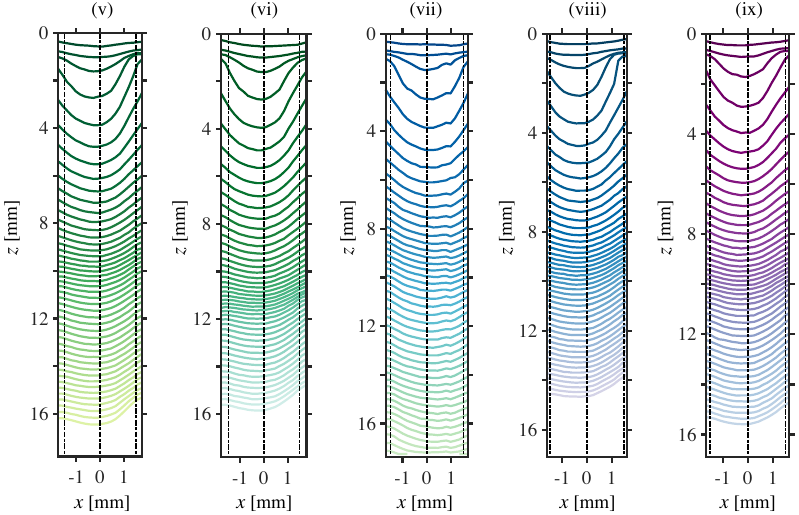}
	\caption{{Propagation of the projected front surface corresponding for experiments (v--ix)]. See caption of figure 3 for details.}}
	\label{fig_frontsb}
\end{figure}

\begin{figure}
	\centering
	\includegraphics[width=0.44\textwidth]{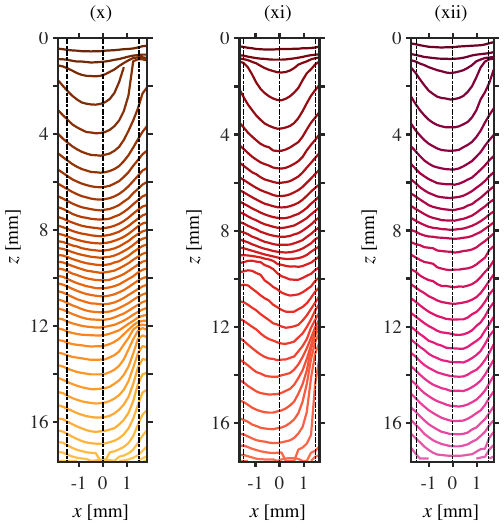}
	\caption{{Propagation of the projected front surface of experiments (x--xii). See caption of figure 3 for details.}}
	\label{fig_frontsc}
\end{figure}

\begin{figure}
	\centering
	\includegraphics[width=1.0\textwidth]{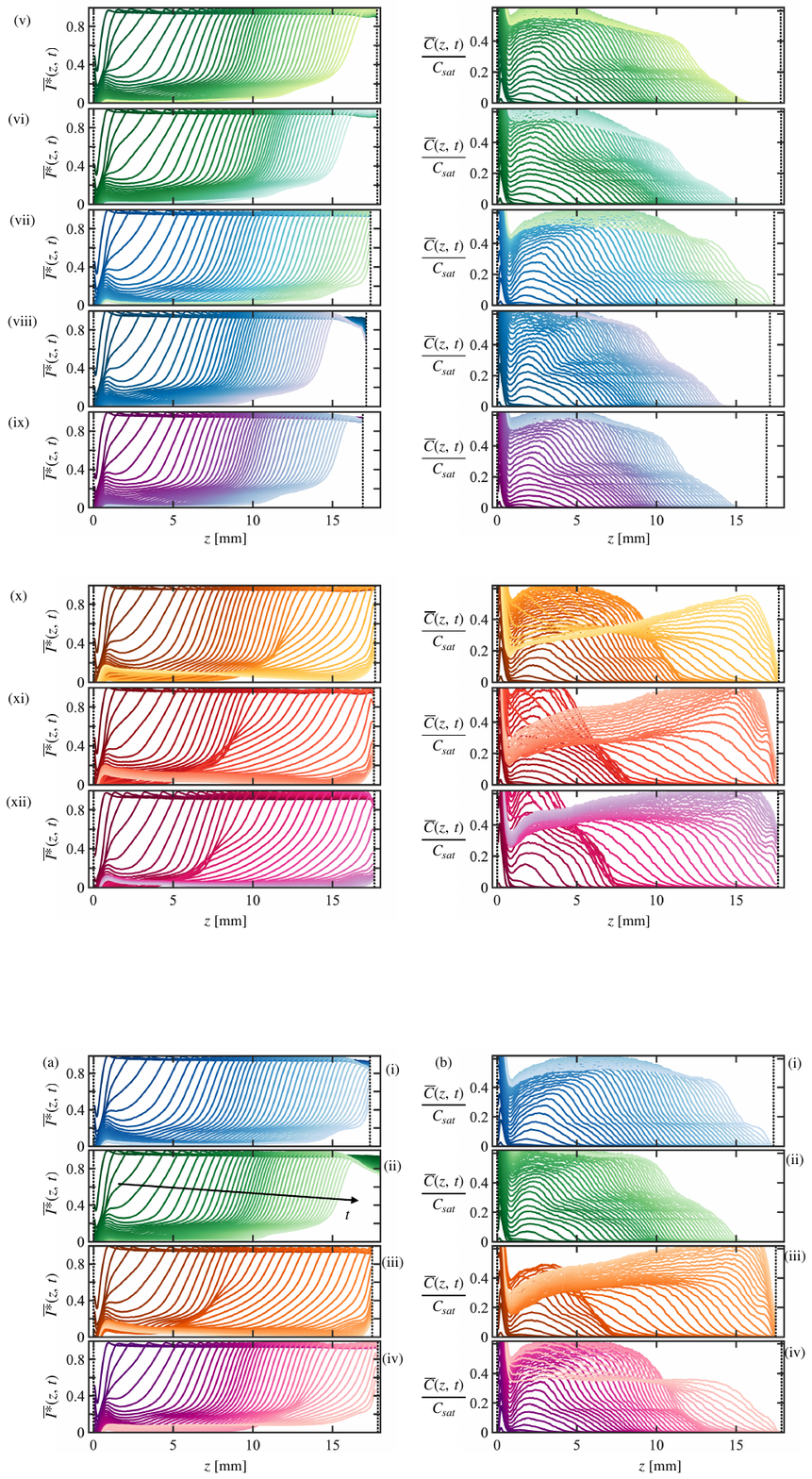}
	\caption{(a) Normalised intensity profiles (left panels)  and corresponding tentative concentration profiles (right panels) for experiments (v--ix). See caption of figures 4 for details.}
	\label{fig_profilesb}
\end{figure}

\begin{figure}
	\centering
	\includegraphics[width=1.0\textwidth]{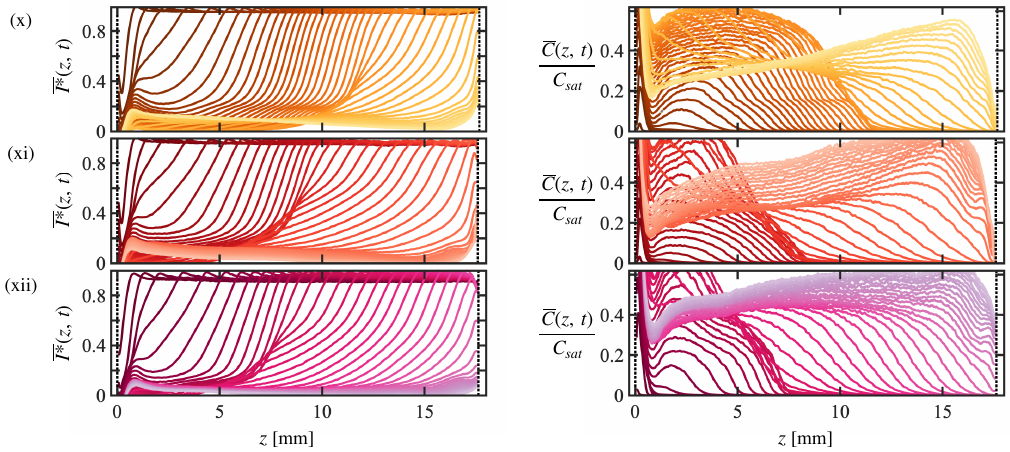}
	\caption{(a) Normalised intensity profiles (left panels)  and corresponding tentative concentration profiles (right panels) for experiments (x--xii). See caption of figures 4 for details.}
	\label{fig_profilesc}
\end{figure}

\begin{figure}
	\centering
	\includegraphics[width=1\textwidth]{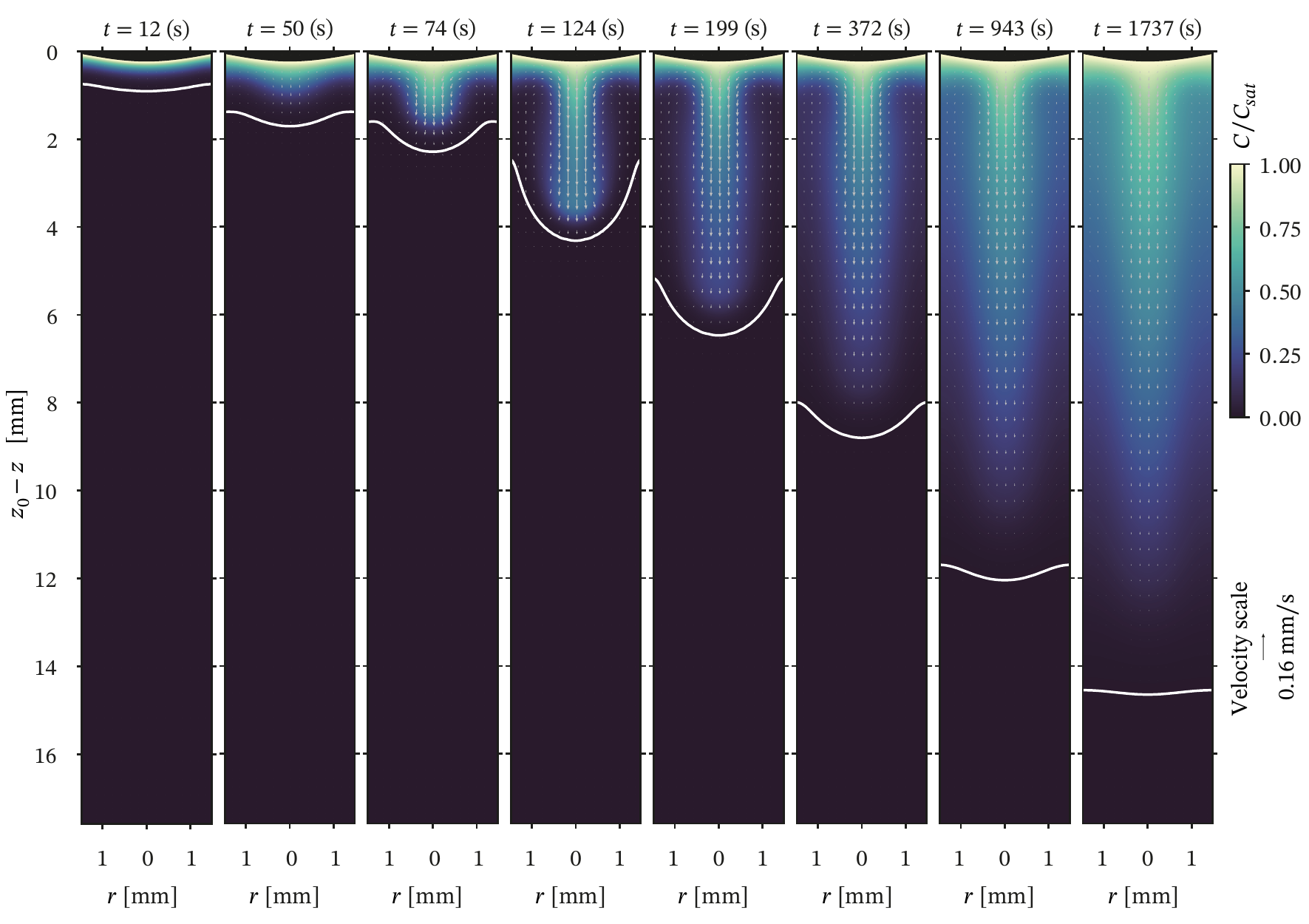}
    \caption{Time evolution of carbon dioxide concentration obtained from numerical simulations for case (B) where the top boundary is modelled as a meniscus interface. The white contour-lines show the front profile associated with $C_f/C_{sat}=0.25\%$. Vectors denote the velocity field, the scaling of which has been provided in the figure.}
    \label{fig_sims_contourf_meniscus}
\end{figure}

\begin{figure}
	\centering
	\includegraphics[width=1.0\textwidth]{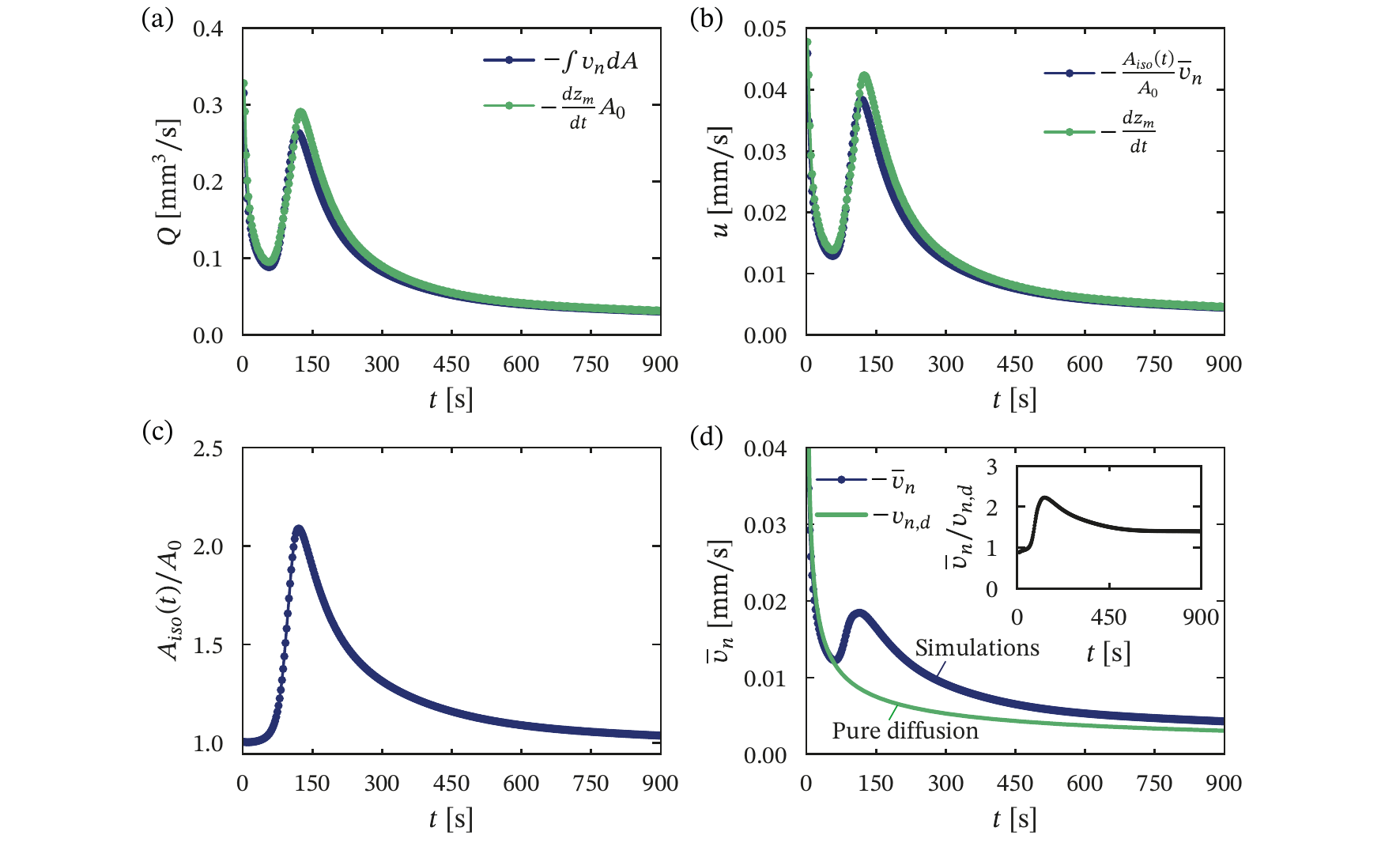}
   \caption{For case (B) (a) Volumetric flux $Q$ of the local propagation velocity $v_n$ across the front iso-surface, compared to the rate of change in the mean volume above the iso-surface, $A_0 dz_m/dt$. (b) Front propagation velocity computed from the time evolution of the cross-sectional average of the 3D iso-surface, $dz_m/dt$, and from the product of the relative surface area and averaged local propagation velocity. $dz_f/dt$ is provided where $z_f$ is the front trajectory obtained from the horizontally-averaged concentration profile in 2D slices, in accordance with the front tracking approach in experiments. (c) Relative surface area of the iso-surface with respect to the cross sectional area of the cylinder, $A_0$. (d) Temporal evolution of the averaged local propagation velocity, $\overline{v}_n$, of the iso-surface. $\overline{v}_n$ for the iso-surface in pure-diffusion problem, compared to the pure diffusive case, $v_{n,d}$, where it equals the front propagation velocity $dz_m/dt= K_f \left[ D(t-t_0) \right]^{-1/2}$, with $K_f$ defined in equation 5. The ratio $\overline{v}_n/v_{n,d}$, is shown in the inset.}
	\label{fig_sims_vn_stats_meniscus}
\end{figure}